\documentclass[12pt]{iopart}
\usepackage{color}
\usepackage{iopams}
\usepackage{setstack}
\usepackage{amstext}
\usepackage{graphicx}
\usepackage{subfigure}
\usepackage{bbold}
\usepackage{ulem}
\usepackage{epsfig}
\usepackage{graphicx}
\usepackage{amssymb}
\usepackage{amsbsy}
\usepackage{amsgen}
\usepackage{amsfonts}    
\usepackage{dcolumn}
\usepackage{amsthm}
\usepackage{mathrsfs}
\usepackage{latexsym}
\usepackage{array}
\usepackage{amstext}
\usepackage{txfonts}

\usepackage{epstopdf} 

\newcommand{\bra}[1]{\langle{#1}\vert}
\newcommand{\ket}[1]{\vert{#1}\rangle}

\newcommand{\inner}[2]{\langle #1 |#2 \rangle}
\newcommand{\abs}[1]{\vert#1\vert}

\begin{document}

\title[]{Robust-to-loss entanglement generation using a quantum plasmonic nanoparticle array}
\author{Changhyoup Lee,$^{1,2}$ Mark Tame,$^{3}$ Changsuk Noh,$^{1}$ James Lim,$^{2,4,5}$\\ Stefan A. Maier,$^{3}$  Jinhyoung Lee,$^{2,4}$ and Dimitris G. Angelakis$^{1,6}$} 
\address{$^1$Centre for Quantum Technologies, National University of Singapore, 3 Science Drive 2, Singapore 117543}
\address{$^2$Department of Physics, Hanyang University, Seoul 133-791, Korea}
\address{$^3$EXSS, The Blackett Laboratory, Imperial College London, Prince Consort Road, SW7 2BW, United Kingdom}
\address{$^4$Center for Macroscopic Quantum Control, Seoul National University, Seoul 151-742, Korea}
\address{$^5$Research Institute for Natural Sciences, Hanyang University, Seoul 133-791, Korea}
\address{$^6$School of Electronic and Computer Engineering,Technical University of Crete, Chania, Crete, Greece 73100}
\ead{changdolli@gmail.com {\rm and} dimitris.angelakis@gmail.com}

\begin{abstract}
We introduce a scheme for generating entanglement between two quantum dots using a plasmonic waveguide made from an array of metal nanoparticles. We show that the scheme is robust to loss, enabling it to work over long distance plasmonic nanoparticle arrays, as well as in the presence of other imperfections such as the detuning of the energy levels of the quantum dots. The scheme represents an alternative strategy to the previously introduced dissipative driven schemes for generating entanglement in plasmonic systems. Here, the entanglement is generated by using dipole-induced interference effects and detection-based postselection. Thus, contrary to the widely held view that loss is major problem for quantum plasmonic systems, we provide a robust-to-loss entanglement generation scheme that could be used as a versatile building block for quantum state engineering and control at the nanoscale.
\end{abstract}




\maketitle

\section{Introduction}
Quantum plasmonics is a rapidly emerging field that offers new opportunities for investigating quantum optics at the nanoscale~\cite{Tame13, Jacob11,DeLeon,Jacob}. Here, novel capabilities in the way the electromagnetic field can be localised~\cite{Reviews1} and manipulated~\cite{Reviews2} open up the prospect of miniaturisation, scalability and strong coherent coupling with single-emitter systems~\cite{Lukin1}, beyond the limits of conventional photonic systems~\cite{Tak}. In particular, with the advancement of nanofabrication and characterisation technologies, metal nanoparticles have been attracting considerable attention as they allow a flexible approach to reaching a high confinement of optical fields~\cite{Nanopart,Krenn99,Nanopart2,Brongersma00}, and it has recently been suggested to use them for building compact on-chip quantum plasmonic networks operating at the nanoscale~\cite{Lee12,Yurke10,Nanonet}. However, Ohmic loss in the metals that support plasmonic excitations is a major obstacle for realising plasmonic quantum information processing and quantum control~\cite{DiMartino12}. This energy dissipation process induces decoherence in the system and limits the performance of a given task. For example, it has been shown that the optimal distance for quantum state transfer in an array of nanoparticles is severely limited~\cite{Lee12}. In order to overcome the problem of loss in plasmonic systems researchers have begun to consider employing various types of gain media~\cite{Berini} or metamaterials~\cite{Kamli}, while trying to keep the high field confinement characteristic offered by plasmonic systems. A different approach has been to tailor the system dynamics such that the steady state of the system yields useful states, the so-called driven-dissipative approach~\cite{diss,Gullans12,diss2}. Despite this important progress in techniques to overcome loss in plasmonic systems, implementing quantum information processing based on an array of metal nanoparticles remains elusive.

In this paper, as a first step towards building quantum plasmonic networks based on metal nanoparticles, we propose an alternative and practical scheme for generating entanglement, between two distant quantum dots (QDs), using the plasmonic modes of a metal nanoparticle array. We show that one can achieve entangled QD states with high fidelities even in the presence of large losses from the metal nanoparticles and inhomogeneous broadenings of the QDs. Robustness against metal loss of the entanglement (or equivalently against an increase in the length of the nanoparticle array) alleviates the length limit of the array, opening up its use for compact nanoscale quantum networks including quantum teleportation~\cite{Bouwmeester97}, quantum communication~\cite{Gisin07}, and quantum repeaters~\cite{Sangouard11}. The plasmonic scenario also provides opportunities in the construction of quantum plasmonic devices on scales far below the diffraction limit, a more compact approach than traditional optical cavities or ion trap networks. The robustness of the entanglement against detunings of the QDs allows the scheme to work even when the QDs have different resonant frequencies, where the operating frequency to be detected is far off-resonant with the QD transitions. This is an important practical consideration. It ultimately stems from the fact that only classical sources of light are required to generate the entanglement between two distant QDs. The scheme we introduce is based on a quantum interference phenomena introduced by Waks and Vu\v{c}kovi\'{c}~\cite{Waks06a} called dipole-induced transparency (DIT). We utilise this interference effect to construct our entangled states and analyse in detail the impact of dissipation and detection efficiency on its stability. While previous studies have investigated entanglement of quantum dots using DIT~\cite{Waks06a}, our analysis of loss in this study goes well beyond that of previous works and allows us to realistically apply DIT to a quantum plasmonic scenario. This provides a more compact setting for the generation of entanglement compared to other approaches, such as photonic crystal cavities, which are an order of magnitude larger in size.


\section{Physical system}

\begin{figure}[t]
\centering
\includegraphics[width=12.5cm]{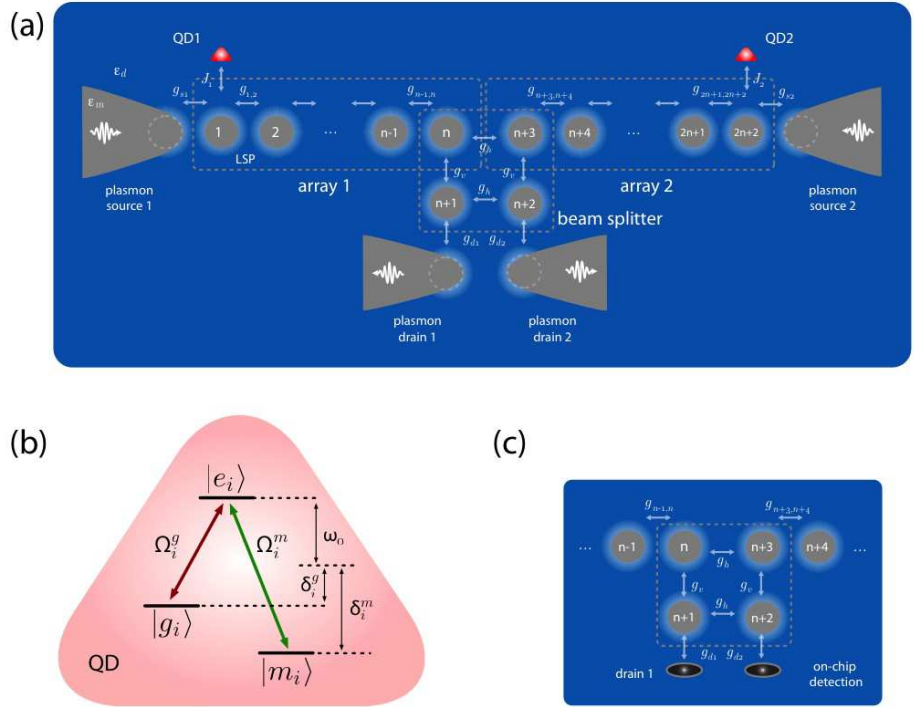}
\caption{(a) Setup for robust-to-loss entanglement generation using an array of metallic nanoparticles. 
The nanoparticle structure can be decomposed into two arms, each consisting of a linear array of nanoparticles, and a special arrangement of nanoparticles in the centre that serves as a plasmonic beam splitter. Two tapered metal nanowire waveguides on the left and right-hand side (source 1 and 2) focus light to the ends of their tips and excite localised surface plasmons on the adjacent nanoparticles. These excitations propagate across the arms and are then mixed at the nanoparticle beam splitter, after which they exit via the other two tapered metal nanowire waveguides (drain 1 and 2). By initialising the quantum dots (QDs) to be in a superposition of ground and excited states, and injecting coherent states of light into both the source nanowires, the system can be tailored so that a detection event at drain 1 signals the generation of entanglement between the QDs. For the detection, an off-chip photon detector is linked to the output signal of drain 1, or alternatively this can be replaced with an on-chip detection unit~\cite{onchipdet} to directly measure the excitation at the $(n+1)$-th nanoparticle (as shown in panel (c)). A detection event at drain 2 (or alternatively on-chip detection linking to the $(n+2)$-th nanoparticle) can also lead to entanglement generation between the QDs. All metallic regions have a frequency-dependent permittivity $\epsilon_{m}(\omega)$, and background dielectric regions have static real and positive permittivity $\epsilon_{d}$, as defined in the text. (b) Energy level diagram for the QDs. Each QD ($i=1,2$) has three states $\ket{g_{i}}$, $\ket{m_{i}}$ and $\ket{e_{i}}$ and has a $\ket{g}$-$\ket{e}$ ($\ket{m}$-$\ket{e}$) transition with frequency $\Omega_{i}^{g~(m)}$, which is detuned by $\delta_{i}^{g~(m)}$ from the natural frequency of the adjacent nanoparticle, $\omega_{0}$.}
\label{setup}
\end{figure}

The physical system for the entanglement generation scheme consists of an array of spherical nanoparticles embedded in a dielectric material, with two distant QDs and four tapered metal nanowires, as depicted in a top-down view in figure~\ref{setup}~(a). All metallic regions have a frequency-dependent permittivity $\epsilon_{m}(\omega)$, and the background dielectric region has a static real and positive permittivity $\epsilon_{d}$. The tapered metal nanowires are connected to adjacent metal nanoparticles and serve as input/output ports. The nanoparticle structure can be decomposed into two arms, each consisting of a linear array of nanoparticles, as considered in Ref.~\cite{Lee12}, which meet in the centre at a special arrangement of four nanoparticles that serves as a plasmonic beam splitter, as discussed in Ref.~\cite{Yurke10}. Each arm of nanoparticles supports electron-charge density oscillations in the longitudinal and transverse directions with respect to the array orientation. For the orientation of the source nanotips shown in figure~\ref{setup}~(a), due to the direction in which the electron charge density oscillates in the nanowires, the nanowire field couples predominantly to the longitudinal oscillation in the nanoparticle. For coupling to the transverse polarisation, we rotate each source nanotip clockwise or counterclockwise by 90$^{\circ}$. Each polarisation direction is maintained through the beam splitter whose corners are bent by 90$^{\circ}$ due to near field nature of the coupling with negligible radiation losses into the far field, provided that only nearest-neighbour interactions are considered~\cite{Maier01,Alu06,Kubota11}. Thus, a longitudinal (transverse) mode from the sources will become a transverse (longitudinal) mode at the drains, and subsequently the drain nanotips should be properly oriented to collect charge-density oscillations in the corresponding direction as shown in figure~\ref{setup}~(a). Alternatively, the out-of-plane transverse mode (perpendicular to the plane) can be used even if the corners at the nanoparticle beam splitter or the junction between the beam splitter and the two arms have arbitrary bending angles. The coupling strength between the nanoparticles depends on their separation distance and the direction of the electron-charge density oscillations \cite{Brongersma00, Jackson75}. Each QD is coupled to the first metal nanoparticle of its respective array (as shown in figure~\ref{setup}~(a)), and assumed to have three internal energy states: a ground state, a long-lived metastable state, and an excited state, which we refer to as $\ket{g}, \ket{m}$ and $\ket{e}$, respectively, as shown in figure~\ref{setup}~(b). The states $\ket{g}$ and $\ket{m}$ represent the qubit states of the QD. The transition between $|g\rangle$ and $|e\rangle$ is assumed to be optically coupled to the adjacent nanoparticle, while the optical transitions connected to $|m\rangle$ are decoupled due to spectral detuning. 
The desired level structure can be realised in a variety of solid-state material systems \cite{level}. 

\subsection{Basic operation of the entanglement generation scheme}

If we take a single arm, say the left-hand side of figure~\ref{setup}~(a), then depending on the state of the QD, both the transmission and reflection coefficients of the metal nanoparticle array exhibit different characteristics. When the QD is in the state $\ket{m}$, {\it i.e.} when it is decoupled from the adjacent nanoparticle due to the  large detuning (see \ref{DIR}.1), the characteristics of the transmission and reflection coefficients of the source nanowire reveal a similar behaviour to those of a single array of nanoparticles (see Ref.~\cite{Lee12}), where the plasmonic field is transmitted across the array when it is on-resonance with one of the eigenmodes. On the other hand, when the QD is in the state $\ket{g}$, {\it i.e.} when it is coupled to the adjacent nanoparticle, the propagating plasmons cannot pass through the adjacent nanoparticle and are completely reflected back into the source nanowire due to a phenomenon called {\it dipole induced reflection} (DIR), which is an alternative form of {\it dipole induced transparency} (DIT)~\cite{Waks06a}. Here, DIR occurs when the operating frequency $\omega$ of the plasmons is resonant with the $\ket{g}$-$\ket{e}$ transition frequency (as shown in \ref{DIR}.1). The entanglement generation scheme then works as follows: Consider that light in the form of plasmon excitations are injected into the system via both of the source nanowires at the same time and that the QDs are prepared in a superposition, $\frac{1}{\sqrt{2}}(\ket{g}_1+\ket{m}_1)\otimes \frac{1}{\sqrt{2}}(\ket{g}_2+\ket{m}_2)$. If a plasmon is detected coming out from drain 1 for given input states at the sources, for example coherent states, then the plasmon came from the cases where either one of QDs is in $\ket{m}$, or both are in $\ket{m}$. A destructive interference of the case $\ket{m}_1\otimes\ket{m}_2$ can be induced by controlling the phases of the initial coherent states, so that only states $\ket{m}_1\otimes \ket{g}_2$ and $\ket{g}_1\otimes\ket{m}_2$ lead to the detection event and consequently a superposed state of $\ket{g}_1\otimes\ket{m}_2$ and $\ket{m}_1\otimes\ket{g}_2$ is generated after the detection, {\it i.e.} $\ket{\psi^-}=\frac{1}{\sqrt{2}}(\ket{m}_1\otimes\ket{g}_2-\ket{g}_1\otimes\ket{m}_2)$, which is an entangled state of the QDs. Likewise, the detection of a plasmon coming from drain 2 can be also used for generating the entangled state $\ket{\psi^{-}}$. This requires different phases of the initial states compared to the case of the detection at drain 1, similar to the case of a bulk beamsplitter which has the transformation $\ket{\alpha}\ket{\beta} \rightarrow \ket{(\alpha+\beta)/\sqrt{2}}\ket{(\alpha-\beta)/\sqrt{2}}$ so the output port is controlled by the input phases. Here, we focus on the case of the detection at drain 1. This is the basic concept for generating entanglement between two QDs using the plasmonic modes of a metal nanoparticle array in the ideal case when there is no loss. Here, we have made use of the characteristics of the transmission and reflection coefficients depending on the internal state of the QDs and an appropriate detection event.

\subsection{System Hamiltonian}

Before describing the entanglement generation scheme in more detail and showing its robustness to loss from the nanoparticles, we first briefly introduce the model to describe the system in figure \ref{setup} (a). We begin with the Hamiltonian for the nanoparticle system which is given by 
\begin{eqnarray}
\hat{H}_{\rm np}=\sum_{i=1}^{2n+2}\hbar \omega_{i} \hat{a}_{i}^\dag \hat{a}_{i} +\sum_{[i,j]}\hbar g_{i,j}\left( \hat{a}^\dag_i \hat{a}_j+\hat{a}^\dag_j \hat{a}_i \right),
\label{Hnp}
\end{eqnarray}
where $\omega_{i}$ is the natural frequency of the field oscillation at the $i$-th nanoparticle, $g_{i,j}$ is the coupling strength between the fields of the $i$-th and $j$-th nanoparticles, and  $[i,j]$ denotes a summation over adjacent neighbours $j$ for a given nanoparticle $i$. The natural frequency $\omega_{i}$ satisfies the Fr\"ohlich criterion ${\rm Re}[\epsilon_{m}(\omega_{i})] =-2 \epsilon_{d}$~\cite{Brongersma00,Maier07}, which considers the nanoparticles to be small enough compared to the operating wavelength such that only dipole-active excitations are important~\cite{Zayats05}. The dielectric function of the metal $\epsilon_{m}(\omega)$ is given by a Drude-Sommerfeld model, which gives rise to a best fit to experimental data at frequencies corresponding to free space wavelengths $\lambda_{0} \gtrsim 350~(530)$ nm for silver (gold)~\cite{Johnson72}. For simplicity we take all nanoparticles to have the same permittivity $\epsilon_{m}$ and radius $R$, thus the local frequencies are set to be equal, $\omega_{i}=\omega_{0}$, for $i=1,\cdots,2n+2$. The operators $\hat{a}^\dag_i$ ($\hat{a}_i$) represent creation (annihilation) operators associated with a dipole-field excitation at the $i$-th nanoparticle, which obey bosonic commutation relations $[\hat{a}_i,\hat{a}^\dag_j]=\delta_{ij}$. Here, a macroscopic quantisation of the fields is used, where the field modes are defined as localised solutions to Maxwell's equations satisfying the boundary conditions of the metal-dielectric interface~\cite{Waks10}. In this case, the electron response in the metal is contained within the dielectric function of the metal, $\epsilon_{m}(\omega)$~\cite{Lee12,Tame08,Elson71}.

The interaction term in equation~(\ref{Hnp}) involves two approximations for the center-to-center distance, $d$, between nanoparticles: a {\it point-dipole} approximation ($3R \leqslant d$), where multipolar interactions are negligible~\cite{Park04}, and a {\it near-field} approximation, $(d \ll \lambda)$, where $\lambda$ is the wavelength of the nanoparticle dipole field~\cite{Yurke10, Krenn99}, where the nearest-neighbour interaction is dominant via the F\"oster fields with a $d^{-3}$ distance dependence~\cite{Maier03}. These two approximations for $d$ are covered simultaneously by a {\it weak-coupling} approximation, where the couplings between nanoparticles, $g_{i,j}$, are much less than the natural frequency, $\omega_{0}$,~\cite{Brongersma00, Yurke10}, {\it i.e.}, $\abs{g_{i,j}}\ll \omega_{0}$ for which we set max$|g_{i,j}| = 0.1 \omega_{0}$ throughout the work. 

The Hamiltonians for the QDs are then given by 
\begin{eqnarray}
\hat{H}_{\rm QD_{1}}^{g~(m)} &= \frac{1}{2} \hbar \Omega_{1}^{g~(m)} \hat{\sigma}_{z,1}^{g~(m)} +\hbar J_{1}^{g~(m)} \left( \hat{a}_{1} \hat{\sigma}^{g~(m) \dagger}_{1} + \hat{a}_{1}^{\dagger} \hat{\sigma}_{1}^{g~(m)} \right)\nonumber\\
\hat{H}_{\rm QD_{2}}^{g~(m)} &= \frac{1}{2} \hbar \Omega_{2}^{g~(m)} \hat{\sigma}_{z,2}^{g~(m)} +\hbar J_{2}^{g~(m)} \left( \hat{a}_{2n+2} \hat{\sigma}^{g~(m) \dagger}_{2} + \hat{a}_{2n+2}^{\dagger} \hat{\sigma}_{2}^{g~(m)} \right),\nonumber
\end{eqnarray}
where $\Omega_{1}^{g~(m)}$ is the transition frequency between $\ket{g}$ and $\ket{e}$ ($\ket{m}$ and $\ket{e}$) of QD$_{1}$ on the lefthand side, $\hat{\sigma}_{1}^{g~(m)}=\ket{g}\bra{e}~~(\ket{m}\bra{e})$, $\hat{\sigma}_{1}^{g~(m) \dagger}=\ket{e}\bra{g}~~(\ket{e}\bra{m})$ and $\hat{\sigma}_{z,1}^{g~(m)}=\ket{e}\bra{e}-\ket{g}\bra{g}~~(\ket{e}\bra{e}-\ket{m}\bra{m})$. The vacuum Rabi frequency of QD$_{1}$ when coupled to the first nanoparticle dependent on the individual directions of the pairwise dipole moments of $\ket{g}$-$\ket{e}$ and $\ket{m}$-$\ket{e}$ transitions of the QD, $J_{1}^{g~(m)}$, is obtained within the rotating-wave approximation and the dipolar approximation, where the distance between the nanoparticle and QD$_{1}$ is assumed to be larger than the radius of the nanoparticle~\cite{Waks10, Trugler08}. Similar denotations and considerations as given above are used for QD$_{2}$ on the righthand side. 

\subsection{Equations of motion for the system}

With the Hamiltonians introduced, the total system including the nanoparticle baths (for modelling loss) and the nanowires (sources and drains) can then be described in terms of Heisenberg equations of motion by using input-output formalism~\cite{Lee12,Collett84,Walls94}. The Heisenberg equations for the lefthand side of the system (from $1$st to ($n+1$)-th nanoparticle and QD$_{1}$) are given by 
\begin{eqnarray}
\fl \frac{d\hat{a}_{1}}{dt} &= -(i \omega_{0} + \frac{g_{\rm s_{1}}}{2}+\frac{\Gamma_{1}}{2}) \hat{a}_{1} - i g_{1,2} \hat{a}_{2} + \sqrt{g_{\rm s_{1}}}\hat{s}_{\rm in,1} + \sqrt{\Gamma_{1}}\hat{b}_{\rm in,1} - i J_{1}^{g~(m)} \hat{\sigma}_{1}^{g~(m)}\label{a1}\\
\fl \frac{d\hat{a}_{j}}{dt} &= -(i \omega_{0} +\frac{\Gamma_{j}}{2}) \hat{a}_{j} - i g_{j,j-1} \hat{a}_{j-1} - i g_{j,j+1} \hat{a}_{j+1} + \sqrt{\Gamma_{j}}\hat{b}_{\rm in,j} ~~~~~~~~ (j=2,3,\cdots,n-1) \label{aj}\\
\fl \frac{d\hat{a}_{n}}{dt} &= -(i \omega_{0} +\frac{\Gamma_{n}}{2}) \hat{a}_{n} - i g_{n,n-1} \hat{a}_{n-1} - i g_{n,n+1}\hat{a}_{n+1} - i g_{n,n+3}\hat{a}_{n+3} + \sqrt{\Gamma_{n}}\hat{b}_{\rm in,n}\label{an}\\
\fl \frac{d\hat{a}_{n+1}}{dt} &= -(i \omega_{0} + \frac{g_{\rm d_{1}}}{2} +\frac{\Gamma_{n+1}}{2}) \hat{a}_{n+1} - i g_{n,n+1}\hat{a}_{n} - i g_{n+1,n+2}\hat{a}_{n+2} + \sqrt{g_{\rm d_{1}}}\hat{d}_{\rm in,1} + \sqrt{\Gamma_{n+1}}\hat{b}_{\rm in,n+1}\label{an+1}\\
\fl \frac{d\hat{\sigma}_{1}^{g~(m)}}{dt}&=-(i\Omega_{1}^{g~(m)} +\frac{\gamma_{1}^{g~(m)}}{2})\hat{\sigma}_{1}^{g~(m)}+i J_{1}^{g~(m)}\hat{a}_{1} \hat{\sigma}_{z,1}^{g~(m)} +\hat{f}_{1}^{g~(m)}.\label{sigma1}
\end{eqnarray}
Similar Heisenberg equations can be written for the righthand side of the system (from ($n+2$)-th to ($2n+2$)-th nanoparticle and QD$_{2}$). The couplings between the nanowires and their adjacent nanoparticles are represented by $g_{\rm s_{1}}$, $g_{\rm s_{2}}$, $g_{\rm d_{1}}$ and $g_{\rm d_{2}}$. They are also assumed to obey the weak-coupling approximation, $|g_{\rm s_{1}}|$, $|g_{\rm s_{2}}|$, $|g_{\rm d_{1}}|$ and $|g_{\rm d_{2}}|$ $\ll \omega_{0}$, as for the interparticle coupling strengths $g_{i,j}$. This is imposed in our model for the nanowires, as the field profiles at the nanowire tips are similar in form to those of the nanoparticles, so that the model is a consistent description \cite{Lee12}. The damping rate $\Gamma$ for each metal nanoparticle depends on its size, and is given by Matthiessen's rule \cite{Brongersma00}: $\Gamma=v_{F}/\lambda_{B} + v_{F}/R$, where $\lambda_{\rm B}$ is the bulk mean-free path of an electron, $v_{F}$ is the velocity at the Fermi surface, and $R$ is the radius. By considering all the nanoparticles to have the same size and permitivity enables us to set $\Gamma_{i}=\Gamma_{0}$ for $i=1,\cdots,2n+2$. The decay rate $\gamma^{g~(m)}$ of a QD from $\ket{e}$ to $\ket{g}$ ($\ket{e}$ to $\ket{m}$) is given by the standard Wigner-Weisskopf spontaneous emission rate of the dipole, where it is assumed that the nanoparticle does not modify it. The operators $\hat{s}_{\rm in,1(2)}$, $\hat{d}_{\rm in,1(2)}$, and $\hat{b}_{\rm in,i}$ represent the input fields of the source nanowires, the drain nanowires, and the $i$-th nanoparticle's bath mode, respectively. The output fields of the nanowires, $\hat{s}_{\rm out,1(2)}$ and $\hat{d}_{\rm out,1(2)}$, and the nanoparticles' baths, $\hat{b}_{\rm out,i}$, are related to the input fields by $\hat{a}_{1(2n+2)}=\frac{1}{\sqrt{g_{\rm s_{1}(s_{2})}}}(\hat{s}_{\rm in,1(2)}+\hat{s}_{\rm out,1(2)})$, $\hat{a}_{n+1(n+2)}=\frac{1}{\sqrt{g_{\rm d_{1}(d_{2})}}}(\hat{d}_{\rm in,1(2)}+\hat{d}_{\rm out,1(2)})$, and $\hat{a}_{i}=\frac{1}{\sqrt{\Gamma_{0}}}(\hat{b}_{\rm in,i}+\hat{b}_{\rm out,i})$, for $i=1,\cdots,2n+2$, respectively. Metal losses in the nanowires are not included in the above Heisenberg equations, but will be included and discussed later. The damping rates of the QDs are assumed to be relatively small compared to all other rates, so that the noise operators, $\hat{f}_{i}^{g~(m)}$ $(i=1,2)$, can be neglected in that they do not significantly affect the quantum coherence of the system. 

\subsection{Scattering matrix for the system}

The entanglement generation scheme works in the weak excitation regime, where the dipole is assumed to be predominantly in the initial state $\ket{g}$ or $\ket{m}$ (the validity of this regime is discussed in~\ref{pulse}, where we consider the pulse profiles of the inputs). In this limit, $\hat{\sigma}_{z}^{g~(m)}$ can be replaced by its mean value, $\langle \hat{\sigma}_{z}^{g~(m)} \rangle \approx -1$, so that the coupled Heisenberg equations of motion give rise to four scattering matrices, one for each of the four internal states of the QDs ($\ket{mm}$, $\ket{gm}$, $\ket{mg}$ and $\ket{gg}$), with the general form
\begin{eqnarray} 
\fl \hat{s}_{\rm in,1}^{\dagger}(\omega)&=t_{\rm s_{1}s_{1}}(\omega) \hat{s}_{\rm out,1}^{\dagger}(\omega)+t_{\rm s_{1}s_{2}}(\omega)\hat{s}_{\rm out,2}^{\dagger}(\omega) +t_{\rm s_{1}d_{1}}(\omega) \hat{d}_{\rm out,1}^{\dagger}(\omega)+t_{\rm s_{1}d_{2}}(\omega)\hat{d}_{\rm out,2}^{\dagger}(\omega)+\hat{f}_{\rm out, s_{1}}^{\dagger}(\omega) \nonumber\\
\fl \hat{s}_{\rm in,2}^{\dagger}(\omega)&=t_{\rm s_{2}s_{1}}(\omega) \hat{s}_{\rm out,1}^{\dagger}(\omega)+t_{\rm s_{2}s_{2}}(\omega)\hat{s}_{\rm out,2}^{\dagger}(\omega) +t_{\rm s_{2}d_{1}}(\omega) \hat{d}_{\rm out,1}^{\dagger}(\omega)+t_{\rm s_{2}d_{2}}(\omega)\hat{d}_{\rm out,2}^{\dagger}(\omega)+\hat{f}_{\rm out, s_{2}}^{\dagger}(\omega),\nonumber\\
\fl &
\label{SM}
\end{eqnarray}
where the noise operators for the nanoparticles are $\hat{f}_{\rm out, s_{1}}^{\dagger}(\omega)=\sum_{i}^{2n+2} t_{\rm s_{1}b_{i}}(\omega)\hat{b}_{\rm out,i}^{\dagger}(\omega)$ and $\hat{f}_{\rm out, s_{2}}^{\dagger}(\omega)=\sum_{i}^{2n+2} t_{\rm s_{2}b_{i}}(\omega)\hat{b}_{\rm out,i}^{\dagger}(\omega)$. The coefficient $t_{\rm \square \lozenge }$ denotes the transition amplitude from an input mode ``$\square$" to an output mode ``$\lozenge$". According to the state of the QDs ($\ket{gg}, \ket{gm}, \ket{mg}$, and $\ket{mm}$), the transition amplitude from ``$\square$" to ``$\lozenge$" when the states of the two QDs are in $\ket{x}\otimes\ket{y}$, for $x,y \in \{ m, g\}$ is represented by the coefficient $t_{\rm \square \lozenge }^{xy}$. For the sake of generality we have defined a dipole operator $\hat{\sigma}_{1(2)}^{m}$ for the dipole state $\ket{m}$, with the $\ket{m}$-$\ket{e}$ transition frequency $\Omega_{1(2)}^{m}$ and damping rate $\gamma_{1(2)}^{m}$. However, in what follows we will treat the system when QD$_{1(2)}$ is in state $\ket{m}$ as if there is no QD$_{1(2)}$, {\it i.e.} $J_{1(2)}^{m}=0$, since it is effectively decoupled from the nanoparticle in this case. For simplicity, we choose all the couplings between nanoparticles in the two arms to be equal, $g_{j,k}=g_{\rm np}$ (at a fixed distance $d$, array orientation, and nanoparticle size \cite{Brongersma00}), and the nanowire to nanoparticle couplings, $g_{\rm s_{1}}=g_{\rm s_{2}}=g_{\rm d_{1}}=g_{\rm s_{2}}=g_{\rm in-out}$. The weak-coupling approximations for these coupling strengths are equivalent to $|g_{\rm np}| \ll \omega_{0}$ and $|g_{\rm in-out}| \ll \omega_{0}$. We impose these by setting max$|g_{\rm np}|=0.1\omega_{0}$ and max$|g_{\rm in-out}|=0.1\omega_{0}$, which are achieved by varying the distance between the nanoparticles, and between the nanowire tips and their adjacent nanoparticles, respectively, for given polarisations of electron-charge density oscillations~\cite{Lee12}. In this work we will mainly focus on the case of  $\Delta\omega=\omega-\omega_{0}=0$, which is approximately the same as the case of using an optical pulse with a small enough bandwidth around $\omega_{0}$ such that all coefficients in equation~(\ref{SM}) are slowly varying. The transitions from the ground (metastable) states to the excited states for the QDs are detuned by $\delta_{1}^{g~(m)}$ and $\delta_{2}^{g~(m)}$ from the resonant frequency $\omega_{0}$ of the nanoparticles, as shown in figure \ref{setup} (b).

From equation~(\ref{SM}), the relation of the operators for the plasmons at resonance, $\omega=\omega_{0}$, can be written as a reduced matrix for each of the four internal states of the QDs to give
\begin{equation}
\left(\begin{array}{c} \hat{s}_{\rm in,1}^{\dagger}(\omega_{0}) \\ \hat{s}_{\rm in,2}^{\dagger}(\omega_{0}) \end{array}\right)
=\left(\begin{array}{cc} 
t_{\rm s_{1}d_{1}}(\omega_{0}) & t_{\rm s_{1}d_{2}}(\omega_{0})  \\
t_{\rm s_{2}d_{1}}(\omega_{0}) & t_{\rm s_{2}d_{2}}(\omega_{0})  \\
\end{array}\right)
\left(\begin{array}{c} \hat{d}_{\rm out,1}^{\dagger}(\omega_{0}) \\ \hat{d}_{\rm out,2}^{\dagger}(\omega_{0}) \end{array}\right),
\label{BS}
\end{equation}
where $| t_{\rm s_{1(2)}d_{1}}(\omega_{0}) |^{2}+| t_{\rm s_{1(2)}d_{2}}(\omega_{0}) |^{2} \leqslant 1$ as the fields are reduced due to the losses from the metal nanoparticles and the QDs, as well as transmissions (or back reflections) to the source nanowire modes $\hat{s}_{\rm out,1(2)}^{\dagger}$. The equality holds only if there is no dissipation and back reflections. A 50/50 beam splitter of lossy nanoparticles is achieved by setting  $g_{n,n+3}=g_{n+1,n+2}=g_{\rm h}=(g_{\rm in-out} + \Gamma_{0})/2$ and $g_{n,n+1}=g_{n+2,n+3}=g_{\rm v}= \sqrt{2} g_{\rm h}$, while taking care of the polarisation-dependence of the couplings between nanoparticles, such that $t_{\rm s_{1}d_{2}}(\omega_{0}) = i t_{\rm s_{1}d_{1}}(\omega_{0}) $ and $t_{\rm s_{2}d_{1}}(\omega_{0}) = i t_{\rm s_{2}d_{2}}(\omega_{0})$ in equation (\ref{BS}) regardless of the internal states of the QDs. Here we assume no cross couplings between nanoparticles at the sites $n$ and $(n+2)$ or $(n+1)$ and $(n+3)$ or $(n-1)$ and $(n+1)$ as they are relatively small compared to $g_{\rm v}$ and $g_{\rm h}$ with a $d^{-3}$ distance dependence~\cite{Maier03}. The coupling values for the 50/50 beam splitter set the transition amplitudes $t_{\rm s_{1}s_{2}}(\omega_{0}) =t_{\rm s_{2}s_{1}}(\omega_{0}) =0$ in equation (\ref{SM}), which implies that any plasmonic field coming from the lefthand (righthand) side source nanowire is not influenced by the state of the QD on the righthand (lefthand) side, $\it i.e.$ we have $t_{\rm s_{1} \lozenge}^{\rm xy}=t_{\rm s_{1} \lozenge}^{\rm xy'}$ and $t_{\rm s_{2} \lozenge}^{\rm xy}=t_{\rm s_{2} \lozenge}^{\rm x'y}$. Note that the configuration of four nanoparticles we use here provides the correct symmetric splitting operation~\cite{Yurke10}.

\begin{figure}[t]
\centering
\includegraphics[width=13cm]{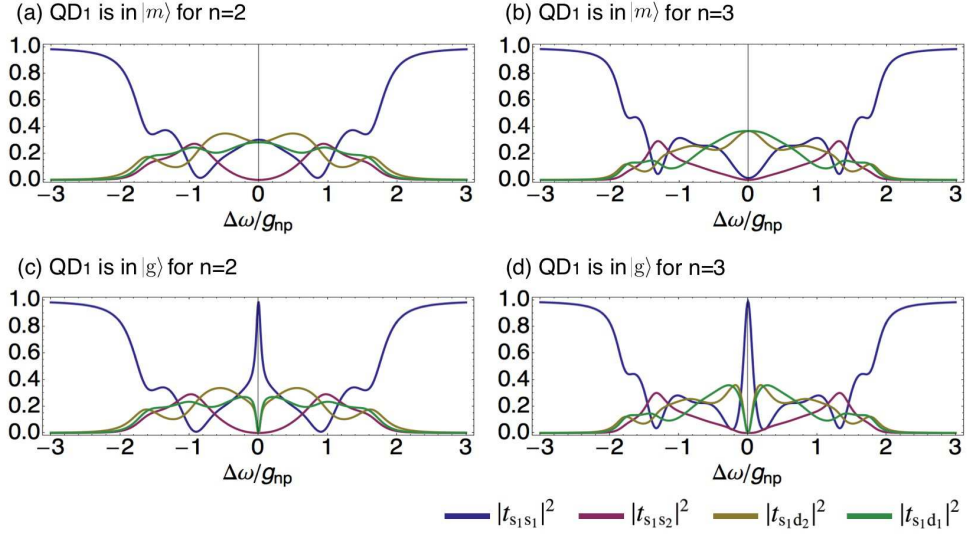}
\caption{Transmission and reflection spectral profiles for plasmons that are injected into the left source nanowire. Left (right) column shows the case of $n=2~(3)$, and the upper (lower) row represents the case when QD$_{1}$ is in $\ket{m}~(\ket{g})$. Regardless of the number of nanoparticles or the state of QD$_{1}$, at resonance ($\Delta \omega=0$) the beam splitter splits the transmitted light field into two halves and blocks any field from reaching the right-hand side arm, so that $\abs{t_{\rm s_{1}d_{1}}}^{2}=\abs{t_{\rm s_{1}d_{2}}}^{2}$ and $\abs{t_{\rm s_{1}s_{2}}}^{2}=0$. In (a) and (b), when QD$_{1}$ is in $\ket{m}$, each coefficient exhibits the characteristics of a basic metal nanoparticle array, where $n=2$ has $\abs{t_{\rm s_{1}s_{1}}}^{2}\neq0$ while $n=3$ has $\abs{t_{\rm s_{1}s_{1}}}^{2}\approx0$  (see main text for details). In (c) and (d), when QD$_{1}$ is in $\ket{g}$, DIR causes $\abs{t_{\rm s_{1}s_{1}}}^{2}\approx1$ for both $n=2$ and $n=3$ when $\Delta \omega=0$.}
\label{spectrum}
\end{figure}

The role of the 50/50 beam splitter in our scheme is to erase the `which-way' information of the two pathways associated with the QD internal states $\ket{gm}$ and $\ket{mg}$. Hence, the amount of energy that leaks into other output modes at the beam splitter is not important as long as we can perfectly erase the which-way information by having the same transition probabilities for the two pathways. Such a lossy beam splitter of nanoparticles is based on the lossless beam splitter proposed by Yurke and Kuang~\cite{Yurke10}. For the case $\Delta\omega \neq 0$, the above choice of coupling values for $g_{\rm h}$ and $g_{\rm v}$ does not guarantee that the nanoparticle configuration performs as a 50/50 beam splitter, {\it i.e.} it leads to a failure in the ability of the beamsplitter to erase the which-way information. This introduces decoherence into the generated entangled state. In figure~\ref{spectrum}, concentrating on the left arm and the beam splitter of nanoparticles, we plot the transmission and reflection spectral profiles for plasmons injected into the left source nanowire, $\abs{t_{\rm s_{1}s_{1}}}^{2}$, $\abs{t_{\rm s_{1}s_{2}}}^{2}$, $\abs{t_{\rm s_{1}d_{1}}}^{2}$ and $\abs{t_{\rm s_{1}d_{2}}}^{2}$ as $\Delta \omega/g_{\rm np}$ is varied, with $g_{\rm in-out}/g_{\rm np}=0.5$, $\Gamma_{0}/g_{\rm np}=0.1$, $J_{1}^{g}/g_{\rm np}=0.3$, and $\gamma_{1}^{g}/g_{\rm np}=0.001$ chosen as an example, when QD$_{1}$ is in $\ket{m}$ or $\ket{g}$. Here, we focus on the case of $n=2$ (the minimum number of nanoparticles in our scheme) and $n=3$ in the left arm as the representative of even or odd number of nanoparticles because the dependence of the transmission and reflection amplitude on the in-out coupling strength $g_{\rm in-out}$ is different for even and odd due to the resonances of the array (as shown in \ref{DIR}.2). For the case of QD$_{1}$ in state $\ket{m}$ (figure~\ref{spectrum}~(a) and (b)), on resonance ($\Delta \omega =0$), the left arm of nanoparticles follows the characteristics of a single array of nanoparticles and just transfers the plasmon field through the array to the beamsplitter where it is split equally. The reason for the large backscattering ($\abs{t_{\rm s_{1}s_{1}}}^{2}$ probability) in the case of $n=2$ (figure~\ref{spectrum}~(a)) is due to the eigenmodes of the array (from the source to each drain) not being on resonance for $\Delta \omega =0$, whereas for $n=3$ (figure~\ref{spectrum}~(b)) there is a resonant eigenmode at $\Delta \omega =0$ for each drain~\cite{Lee12}. For $\Delta \omega \neq 0$ the various eigenmodes of the system come into resonance at particular frequencies of the input field and cause the oscillatory behaviour seen on either side of $\Delta \omega=0$. On the other hand, for the case of QD$_{1}$ in state $\ket{g}$ (figure~\ref{spectrum}~(c) and (d)), the first nanoparticle is coupled to QD$_{1}$ and DIR occurs (as shown in \ref{DIR}), regardless of the resonant eigenmodes in the nanoparticle system. Thus in both figure~\ref{spectrum}~(c) and (d) at $\Delta \omega =0$ the backscattered probability $\abs{t_{\rm s_{1}s_{1}}}^{2}$ goes to unity. Away from $\Delta \omega =0$ the transition probabilities follow the same behaviour as for the case of QD$_1$ in state $\ket{m}$, as the QD becomes off resonant and so cannot induce any dipole-based interference effect.


\section{Entanglement generation scheme}
In this section, we first consider a weak input classical source to delineate the basic idea of how the scheme works and then describe a more general scenario.

\subsection{Weak field injection}

To generate entanglement between the QDs, both are initialised to be in an equal superposition of qubit states $\ket{g}$ and $\ket{m}$, {\it i.e.} $\ket{+}=\frac{1}{\sqrt{2}}(\ket{g}+\ket{m})$. This can be achieved by first driving the QDs into the lowest-energy state and then rotating them by either a direct $\pi/2$ transition, or a Raman transition \cite{Stievater01}, depending on the specifics of the QD internal energy level structure. After initialising the QDs, a coherent field $\ket{\alpha}$ with frequency $\omega_{0}$ is injected into the left source, $\hat{s}_{\rm in,1}^{\dagger}$, and simultaneously another coherent field $\ket{\beta}$ with same polarisation and frequency as $\ket{\alpha}$ is injected into the right source, $\hat{s}_{\rm in,2}^{\dagger}$. The initial state of the whole system is then
\begin{eqnarray}
\ket{\Psi_{i}}=\frac{1}{2}(\ket{g}_{1}+\ket{m}_{1}) (\ket{g}_{2}+\ket{m}_{2})  \ket{\alpha}_{s_{1}} \ket{\beta}_{s_{2}}  \ket{0}_{d_{1}} \ket{0}_{d_{2}}.\nonumber
\end{eqnarray}
The input plasmon fields interact with the QDs only if they are in the state $\ket{g}$. In the case that one (or both) of the QDs is in the state $\ket{m}$, a field is transferred along one (or both) of the arms, then mixed at the beam splitter and goes out through one of the drain nanowires, $\hat{d}_{\rm out,1(2)}^{\dagger}$. Entanglement can be generated by the detection of an excitation in drain 1, which could have come from either the left or right arm, when the QDs were in $\ket{mg}$ and $\ket{gm}$ respectively (the case when both arms supply an excitation, with the QDs in $\ket{mm}$, is suppressed as described below). Thus, the nanowire mode $\hat{d}_{\rm out,1}^{\dagger}$ is connected to a detector in order to postselect the successful case when the which-way information has been erased and the QDs have been entangled as a consequence. Alternatively, an on-chip detector could be directly coupled to the $(n+1)$-th nanoparticle so that the drain nanowire is not needed. Note that the total plasmonic system only requires a classical source of light in order to generate entanglement.

To see how the scheme works, it is useful to consider sufficiently weak coherent states ($\alpha\ll1, \beta\ll1$) when there is no loss present. We may expand $\ket{\Psi_{i}}$ to first order in the plasmonic excitation number and approximately drop all other output modes. Then, as the scheme is based on the detection of an excitation we can drop the vacuum terms so that $\ket{\Psi_{i}} \approx \frac{1}{2}(\ket{g}_{1}+\ket{m}_{1}) (\ket{g}_{2}+\ket{m}_{2}) (\alpha\ket{1}_{s_{1}} +\beta \ket{1}_{s_{2}})$. Perfect DIR leads to $t_{\rm s_{1} d_{1}}^{gg}=t_{\rm s_{1} d_{1}}^{gm}=0$ and $t_{\rm s_{2} d_{1}}^{gg}=t_{\rm s_{2} d_{1}}^{mg}=0$, and using equation~(\ref{SM}) for each of the internal states of the QDs, together with a postselection of the state when an excitation is present in drain 1, we have the unnormalised state of the QDs as 
\begin{equation}
\ket{\Psi}_{\rm QDs} = \frac{1}{2} \Big\{ (\alpha t_{\rm s_{1}d_{1}}^{mm} +\beta t_{\rm s_{2}d_{1}}^{mm})\ket{mm}+\beta t_{\rm s_{2}d_{1}}^{gm}\ket{gm}+ \alpha t_{\rm s_{1}d_{1}}^{mg} \ket{mg} \Big\}.
\label{PsiQDs}
\end{equation} 
Here, the beam splitter enables $t_{\rm s_{1}d_{1}}^{mm} = t_{\rm s_{1}d_{1}}^{mg}$ and $t_{\rm s_{2}d_{1}}^{mm} = t_{\rm s_{2}d_{1}}^{gm}$, {\it i.e.} the plasmon injected in each source is not influenced by the QD in the opposite arm. If the phase of the amplitude $\beta$ is chosen such that $\alpha t_{\rm s_{1}d_{1}}^{mm} +\beta t_{\rm s_{2}d_{1}}^{mm} =0$, we have a normalised ideal entangled state as $\ket{\Psi}_{\rm QDs}=\frac{1}{\sqrt{2}}( \ket{mg}-\ket{gm})\equiv \ket{\psi^{-}}$ with probability $\abs{ \alpha t_{\rm s_{1}d_{1}}^{mg} }^{2}$. This is the basic idea of the scheme for the limiting case of weak $\alpha$ and $\beta$.

\subsection{Arbitrary field injection and loss}

We now turn to the more general scenario, starting with the initial state $\ket{\Psi_{i}}$ without any restrictions on $\alpha$ and $\beta$. In this case the detection of excitations at drain 1 is modelled by the projection operator $\mathrm{P}_{\rm d_{1}}=\sum_{n=1}^{\infty}{\ket{n}_{\rm d_{1}}\bra{n}}= \mathbb{I} - \ket{0}_{\rm d_{1}}\bra{0}$ that projects the state of the system onto a subspace containing at least one excitation in mode $\hat{d}_{\rm out,1}^{\dagger}$. This projection models the measurement performed by an ideal non-photon number resolving detector, which registers a detection event as long as there is at least one excitation at the detector. Later we will also investigate the effects of detection inefficiency in the scheme. As a result of lifting the restriction on the input field amplitudes and the accuracy of the detection, the state of the QDs becomes mixed and can be written as 
\begin{equation}
\rho_{\rm QDs}=\frac{\mathrm{Tr}_{(\mathrm{all~fields})}[\mathrm{P}_{\rm d_{1}}\ket{\Psi_{f}}\bra{\Psi_{f}}\mathrm{P}_{\rm d_{1}}]}{\mathrm{Tr}_{(\mathrm{QDs})}[\mathrm{Tr}_{(\mathrm{all~fields})}[\mathrm{P}_{\rm d_{1}}\ket{\Psi_{f}}\bra{\Psi_{f}}\mathrm{P}_{\rm d_{1}}]]}.  \label{genout}
\end{equation}
The final state of the output fields, $\ket{\Psi_{f}}$, is found by transforming $\ket{\Psi_{i}}$ according to the four sets of scattering matrices from equation (\ref{SM}) depending on the state of the QDs. It is written as
\begin{eqnarray}
\ket{\Psi_{f}} =\frac{1}{2} \Big( \ket{gg}\ket{\psi^{gg}}_{\rm fields}+\ket{mm}\ket{\psi^{mm}}_{\rm fields} + \ket{gm}\ket{\psi^{gm}}_{\rm fields}+ \ket{mg}\ket{\psi^{mg}}_{\rm fields} \Big),
\label{final state}
\end{eqnarray}
where the state of the fields at the output modes is represented by a product of coherent states
\begin{eqnarray}
\ket{\psi^{xy}}_{\rm fields}=\ket{\xi_{1}^{xy}}_{\rm s_{1}} \ket{\xi_{2}^{xy}}_{\rm s_{2}} \ket{\mu_{1}^{xy}}_{\rm d_{1}} \ket{\mu_{2}^{xy}}_{\rm d_{2}} \otimes_{j=1}^{2n+2} \ket{\chi_{j}^{xy}}_{\rm b_{j}},
\end{eqnarray}
with amplitudes $\xi_{j}^{xy}=\alpha t_{\rm s_{1}s_{j}}^{xy}+\beta t_{\rm s_{2}s_{j}}^{xy}$ $(j=1,2)$ at the source nanowires, $\mu_{j}^{xy}=\alpha t_{\rm s_{1}d_{j}}^{xy}+\beta t_{\rm s_{2}d_{j}}^{xy}$ $(j=1,2)$ at the drain nanowires, and $\chi_{j}^{xy}=\alpha t_{\rm s_{1}b_{j}}^{xy}+\beta t_{\rm s_{2}b_{j}}^{xy}$ $(j=1,\cdots,2n+2)$ for the nanoparticle bath modes. Each coherent field consists of an interference of two pathways, one from source 1 and the other from source 2 with respective transition amplitudes. Throughout this work, we impose the matching condition $\alpha t_{\rm s_{1}d_{1}}^{mm} +\beta t_{\rm s_{2}d_{1}}^{mm} =0$ in order to exclude the possibility of the case of $\ket{mm}$ leading to a detection, {\it i.e.} the amplitude for the coherent state in drain 1 is $\mu_{1}^{mm}=0$. As mentioned previously, this can be achieved by properly adjusting the amplitude $\beta$ of the coherent field injected into source 2. In order to measure how well the QDs are entangled, we employ the fidelity as a figure of merit, defined as the overlap between the desired final state, $\ket{\psi^-}$, and the actual final state of the system, $\rho_{\rm QDs}$, and given by $F=\bra{\psi^-}\,\rho_{\rm QDs}\,\ket{\psi^-}$. The fidelity allows us to quantify how close the final state is to the desired state, and at the same time provides a lower bound on the concurrence, $C$, a typical entanglement measure for two qubits, {\it i.e.} $C\geqslant {\rm max}(0,2F-1)$. Such a lower bound implies that entanglement is always found when the fidelity is greater than $1/2$ \cite{Vert,Bell12}. The fidelity is given as $F=\bra{\psi^{-}} \rho_{\rm QDs} \ket{\psi^{-}}=\frac{1}{2}(\rho_{22}+\rho_{33}-\rho_{23}-\rho_{32})$, where $\rho_{pq}$ is the entry in the $p$-th row and the $q$-th column of the matrix for $\rho_{\rm QDs}$ that is spanned by the basis $\{ \ket{gg}, \ket{gm}, \ket{mg}$, $\ket{mm} \}$. The entries are given by  
\begin{eqnarray}
&&\fl \rho_{22} = \frac{1}{4\eta} \Big(1-\inner{\mu_{1}^{gm}}{0}_{\rm d_{1}}\inner{0}{\mu_{1}^{gm}}_{\rm d_{1}} \Big), ~~~~\rho_{33} = \frac{1}{4\eta} \Big(1-\inner{\mu_{1}^{mg}}{0}_{\rm d_{1}}\inner{0}{\mu_{1}^{mg}}_{\rm d_{1}} \Big) \label{rhos} \\
&&\fl \rho_{23} = \frac{1}{4\eta} \Big(\inner{\mu_{1}^{gm}}{\mu_{1}^{mg}}_{\rm d_{1}}-\inner{\mu_{1}^{gm}}{0}_{\rm d_{1}}\inner{0}{\mu_{1}^{mg}}_{\rm d_{1}} \Big) \times \nonumber \\
&& \qquad \qquad\inner{\mu_{2}^{gm}}{\mu_{2}^{mg}}_{\rm d_{2}} \inner{\xi_{1}^{gm}}{\xi_{1}^{mg}}_{\rm s_{1}} \inner{\xi_{2}^{gm}}{\xi_{2}^{mg}}_{\rm s_{2}} \otimes_{j=1}^{2n+2} \inner{\chi_{j}^{gm}}{\chi_{j}^{mg}}_{\rm b_{j}} =  \rho_{32}^{*}. \nonumber
\end{eqnarray}
The efficiency (or success probability for generating an entangled state) is the probability to detect photons at drain 1 and is given by the denominator of equation (\ref{genout}) as
\begin{eqnarray}
\eta&=&\mathrm{Tr}_{(\mathrm{QDs})}[\mathrm{Tr}_{(\mathrm{all~fields})}[\mathrm{P}_{\rm d_{1}} \ket{\Psi_{f}}\bra{\Psi_{f}}\mathrm{P}_{\rm d_{1}}]] \nonumber\\
&=&1-\frac{1}{4}\Big[\abs{\inner{0}{\mu_{1}^{gg}}}^{2}+\abs{\inner{0}{\mu_{1}^{mm}}}^{2}+\abs{\inner{0}{\mu_{1}^{gm}}}^{2}+\abs{\inner{0}{\mu_{1}^{mg}}}^{2}\Big].
\label{efficiency}
\end{eqnarray}
Thus, by inspecting equation (\ref{rhos}), a high fidelity is obtained in the limiting scenario where we have $\mu_{1}^{gm}=-\mu_{1}^{mg}$ with all the other field modes for the case when the QDs are in $\ket{gm}$ approximately equal to the field modes for the case when the QDs are in $\ket{mg}$, so that they are effectively factored out in equation~(\ref{final state}). In this case, the state of the total system becomes $\ket{\Psi_{f}} = (\mu_{1}^{gm}\ket{gm}+\mu_{1}^{mg}\ket{mg})\otimes\ket{\rm all~fields}$.
\begin{figure}[t]
\centering
\includegraphics[width=12cm]{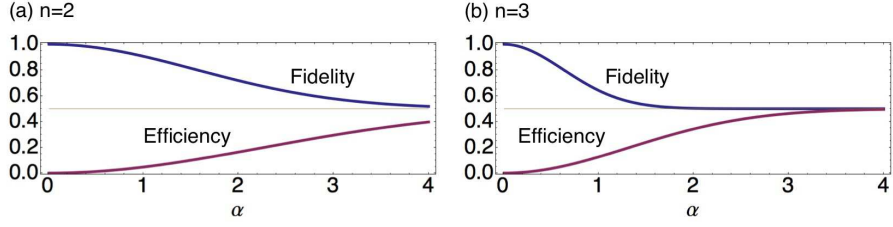}
\caption{Fidelity of the state of the QDs, $\rho_{\rm QDs}$, with respect to the maximally entangled state $\ket{\psi^{-}}$ at resonance $\Delta\omega=0$ for the coherent fields injected into the sources. Here, the magnitude of the field of the injected states, $\alpha$ (with $\beta=i\alpha$), is varied for $n=2$ and $n=3$. As an example, we have used the system parameters $g_{\rm in-out}/g_{\rm np}=0.5$, $J_{1}^{g}/g_{\rm np}=J_{2}^{g}/g_{\rm np}=0.3$, $\gamma_{1}^{g}/g_{\rm np}=\gamma_{2}^{g}/g_{\rm np}=0.001$ and $\Gamma_{0}/g_{\rm np}=0.1$. When $\alpha$ is increased, the fidelities drop and asymptotically approach $0.5$, whereas the efficiencies go up and asymptotically approach $0.5$. The gradients of the fidelities and efficiencies can be changed depending on the system parameters we have set. These affect the performance of the scheme, as shown in figure \ref{fidelity loss}, where we vary $g_{\rm in-out}/g_{\rm np}$ and $\Gamma_{0}/g_{\rm np}$.}
\label{fidelity alpha}
\end{figure}
However, this limiting scenario is hard to achieve for arbitrary fields injected into the system as higher-order photon number contributions act as a decoherence mechanism on the final state. The impact of higher-order photon number contributions on the value of the fidelity and efficiency are shown figure \ref{fidelity alpha}. Here, a trade-off between fidelity and efficiency can be seen as $\alpha$ is increased (note: $\beta=i\alpha$ by the matching condition $\alpha t_{\rm s_{1}d_{1}}^{mm} +\beta t_{\rm s_{2}d_{1}}^{mm} =0$)~\cite{Sridharan08}. The fidelity decays quickly with increasing $\alpha$ due to the presence of higher-order excitations in the plasmonic system. It asymptotically approaches $0.5$ in the limit $\alpha \gg 1$, indicating that the higher-order excitation contributions have completely decohered the state of the QDs. On the other hand, the fidelity reaches close to one when $\alpha \ll 1$, which indicates that an ideal entangled state is generated in the limit of the weak field injection case, as described in the previous subsection. Note that higher-order excitations are inevitable but not a practical problem as the amplitude $\alpha$ is easily controllable, so that we can decrease the magnitude of $\alpha$ as low as required (or increased) for a reasonable fidelity and efficiency.

\section{Results}

Having described the entanglement generation scheme and general trends for the fidelity and efficiency we now go into the details of the robustness of the fidelity to loss from the metal nanoparticles, the length of the arms, and inhomogneous broadenings of the QDs. We also examine the efficiency of generating entanglement in these circumstances.  We will show the main results in this section, and explain more detailed mathematics in the Appendices. Our analysis of entanglement generation is classified into two groups, each according to the number of nanoparticles in the arms: even or odd. This is because the dependence of the transmission amplitude on the in-out coupling strength $g_{\rm in-out}$ is different for even and odd due to the resonances of the array (as shown in \ref{DIR}.2). We have also chosen to plot all parameters in units of the nanoparticle coupling $g_{\rm np}$ so that the plots are independent of $g_{\rm np}$, as long as $g_{\rm np} \ll \omega_{0}$ is satisfied (weak-coupling approximation). Note also that max($g_{\rm in-out}/g_{\rm np}^{\rm max}$)=1 must be imposed, where $g_{\rm np}^{\rm max}=0.1\omega_{0}$, otherwise we move away from the weak-coupling regime for the sources and drains. In other words, the rescaled couplings $g_{\rm in-out}/g_{\rm np}$ can in principle go higher than 1, but the value for $g_{\rm np}$ must be lower than $0.1 \omega_{0}$ to compensate, so that we are still in the weak-coupling regime. All calculations are done for resonant input fields, $\Delta\omega=\omega-\omega_{0}=0$ (the case of pulsed input fields is investigated in~\ref{pulse}). We also set the QD-nanoparticle couplings as $J_{1}^{g}/g_{\rm np}=J_{2}^{g}/g_{\rm np}=0.3$, and the QD decay rates $\gamma_{1}^{g}/g_{\rm np}=\gamma_{2}^{g}/g_{\rm np}=0.001$. Other parameters of the system are varied in our analysis.

\subsection{Robust fidelity against metal loss}

\begin{figure}[t]
\centering
\includegraphics[width=12cm]{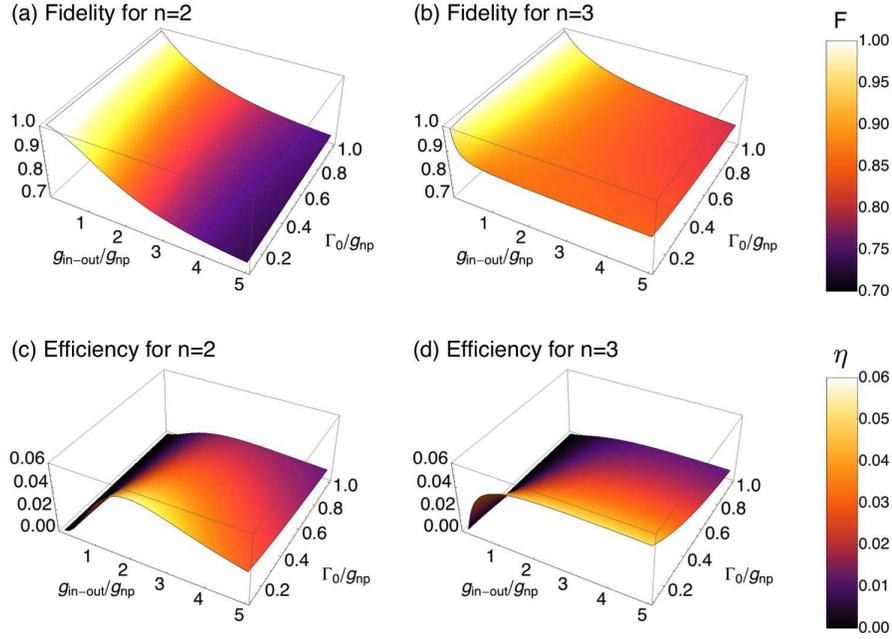}
\caption{Fidelity of generated QD state, $\rho_{\rm QDs}$, with respect to the maximally entangled state $\ket{\psi^{-}}$ when the input fields are on resonance ($\Delta\omega=0$) as the in-out coupling strength $g_{\rm in-out}/g_{\rm np}$ and the amount of metal loss $\Gamma_{0}/g_{\rm np}$ are varied, for $n=2$ and $n=3$. Left (right) hand column shows the case of $n=2~(3)$, and the upper (lower) row shows the fidelity (efficiency).}
\label{fidelity loss}
\end{figure}

We first examine the fidelity and efficiency of our scheme with respect to loss in the metal nanoparticles. For this purpose, we choose the cases of $n=2$ and $n=3$ as representatives of even and odd numbers of nanoparticles in each arm. Figure~\ref{fidelity loss} shows the fidelity and efficiency as the nanoparticle loss $\Gamma_{0}/g_{\rm np}$ increases and the in-out coupling strength $g_{\rm in-out}/g_{\rm np}$ is varied for $\alpha = 0.5$ (with $\beta=i\alpha$). Here, we have assumed that the $\ket{g}$-$\ket{e}$ transition of each QD is resonant with its adjacent nanoparticle such that $\delta_{1(2)}^{g}=0$, while $\delta_{1(2)}^{m}$ is chosen for the QD $\ket{m}$-$\ket{e}$ transition to be completely decoupled from the dynamics.

In figure~\ref{fidelity loss} (a) and (b), we show that the fidelity varies only slightly with increasing $\Gamma_{0}/g_{\rm np}$, implying a remarkable robustness against metal loss. Such a robustness in fidelity can be understood from the observation that the reflection coefficient of each arm predominantly determines the fidelity for the case of resonant QDs ($\delta_{1}^{g}=\delta_{2}^{g}=0$), as we explain in detail in~\ref{detail}. This indicates that the fidelity mostly depends on the behaviour of the reflection coefficient of each arm with metal loss $\Gamma_{0}/g_{\rm np}$ and $g_{\rm in-out}/g_{\rm np}$; the reflection coefficient varies only slightly with $\Gamma_{0}/g_{\rm np}$ since the energy never enters the array, and has a monotonic trend with $g_{\rm in-out}/g_{\rm np}$ (see~\ref{DIR}), resulting in robustness of the fidelity against metal loss and a general decrease in the fidelity with respect to $g_{\rm in-out}/g_{\rm np}$, respectively. Contrary to the case of the fidelity, we show that the efficiency of generating entanglement is decreased with $\Gamma_{0}/g_{\rm np}$ in figure~\ref{fidelity loss} (c) and (d). This is because the efficiency is determined by how much energy is transferred from the sources to the drains, indicating that the efficiency is mostly related to the transmission coefficient of each arm. In contrast to the reflection, the transmission coefficient is more sensitive to metal loss, and has different dependence on $g_{\rm in-out}/g_{\rm np}$ (as shown in more detail in~\ref{DIR}).

In this subsection, we chose $\alpha=0.5$ as an example. However, if we choose a larger value of $\alpha$ then higher-order excitations will reduce the fidelity and increase the efficiency on average in figure \ref{fidelity loss}, as already shown in the previous subsection. Note that this is not a practical problem as the amplitude $\alpha$ is easily controllable, so that it can be decreased as low as required (or increased) for a reasonable fidelity and efficiency. Thus, using coherent states as initial states is a key merit of the scheme, which allows us to use a metal nanoparticle array supporting lossy plasmonic modes.

\subsection{Robust fidelity against the length of a nanoparticle array}

\begin{figure}[t]
\centering
\includegraphics[width=12cm]{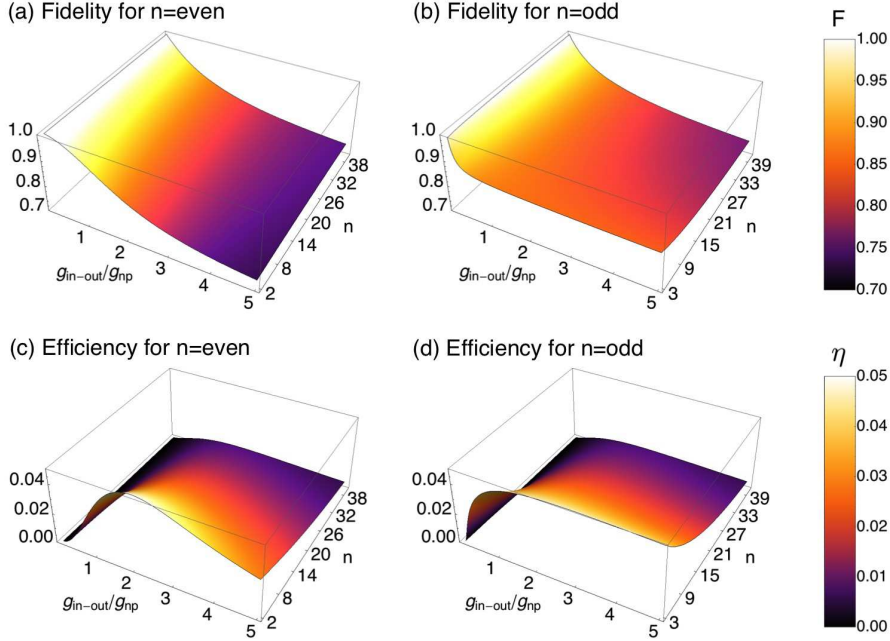}
\caption{Fidelity of generated QD state, $\rho_{\rm QDs}$, with respect to the maximally entangled state $\ket{\psi^{-}}$ when the input fields are on resonance ($\Delta\omega=0$) as the in-out coupling strength $g_{\rm in-out}/g_{\rm np}$ is varied and the number of nanoparticles $n$ is increased for a set amount of metal loss, $\Gamma_{0}/g_{\rm np}=0.1$. Left (right) column presents the case of even (odd) $n$, and the upper (lower) row presents the fidelity and efficiency respectively.}
\label{fidelity np}
\end{figure}

In terms of the total amount of loss present in the entire system, the effect of increasing the loss of each nanoparticle, for a fixed number of nanoparticles, can be regarded as equivalent to increasing the number of nanoparticles in each arm for a fixed amount of loss. In figure~\ref{fidelity np}, for a fixed amount of metal loss, $\Gamma_{0}/g_{\rm np}=0.1$, and input amplitude $\alpha=0.5$, the fidelity and efficiency are shown as $g_{\rm in-out}/g_{\rm np}$ is varied with increasing number of nanoparticles. As expected, the behaviour of the fidelity and the efficiency as the length of nanoparticle array is increased is very similar to the behaviour seen in figure~\ref{fidelity loss}. It is quite remarkable that one can achieve a robust high fidelity (and thus entanglement) over a reasonably long array of metal nanoparticles, even though it has so far been believed that loss limits the length of an array of nanoparticles for quantum information processing purposes. 

Furthermore, we have checked that the number of nanoparticles can be increased arbitrarily with the fidelity staying consistently above 0.8 at the maximum efficiency optimized over $g_{\rm in-out}/g_{\rm np}$ for $\alpha=0.5$ but that at the same time this maximum efficiency asymptotically approaches zero as the loss becomes much more dominant in the overall dynamics. Therefore the overall performance is limited by the repetition rate of the pulses used for entangling. Meanwhile, $g_{\rm np}$ can be arbitrarily changed as long as the value of $g_{\rm in-out}/g_{\rm np}$ is kept within the necessary limits, but if the distance between nanoparticles increases too much ($g_{\rm np}$ decreases), then their coupling becomes so weak that the radiative damping rate becomes important~\cite{Brongersma00}. According to a typical example of a nanoparticle array \cite{Brongersma00}, where nanoparticles with radius of $25$nm are separated by a center-to-center distance $75$nm and our approximations are satisfied, the size of the total array is about $n\times150$nm (for example, if $n=40$, the total size is about $6\mu$m).

\subsection{Robust fidelity against detunings of the QDs}

\begin{figure}[t]
\centering
\includegraphics[width=13cm]{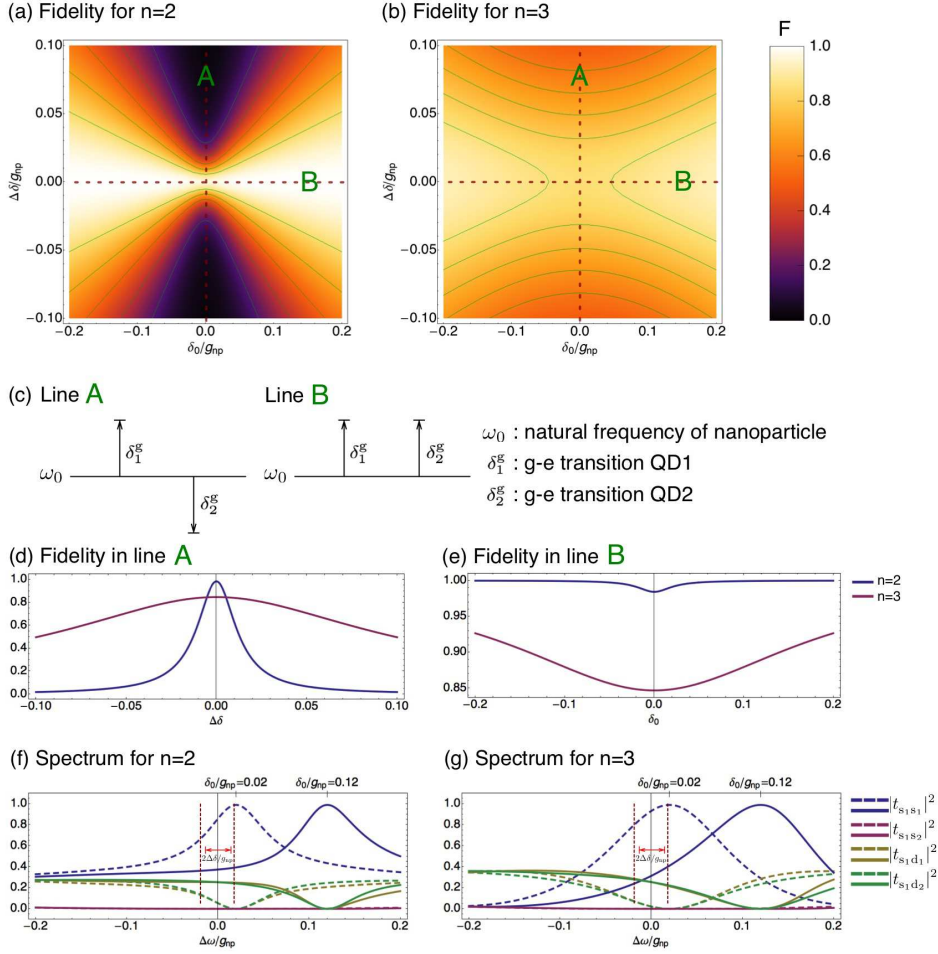}
\caption{Fidelity of generated QD state, $\rho_{\rm QDs}$, with respect to the maximally entangled state $\ket{\psi^{-}}$ when the QDs are detuned. In (a) and (b) the input fields are on resonance ($\Delta\omega=0$) as the average detuning $\delta_{0}/g_{\rm np}$ and the difference in detuning $\Delta\delta/g_{\rm np}$ of the QDs is varied for $n=2$ and $n=3$, respectively. Two extreme regions are labelled: line A is when $\delta_{0}/g_{\rm np}=0$ and $\Delta\delta/g_{\rm np}\neq0$, and line B is when $\delta_{0}/g_{\rm np}\neq0$ and $\Delta\delta/g_{\rm np}=0$. The corresponding energy level diagrams are shown in (c). In (d) and (e) the fidelities in the lines A and B are shown, respectively, for both $n=2$ and $n=3$. In (f) and (g) the reason why the fidelity is more robust against a change in $\Delta\delta/g_{\rm np}$ the more detuned $\delta_{0}/g_{\rm np}$ is from $\omega_{0}$ can be seen from the behaviour of the transmission and reflection spectrums for two separate values of $\delta_{0}/g_{\rm np}$ (solid and dashed) and a fixed value of $\Delta \delta/g_{\rm np}$}
\label{detunings}
\end{figure}

A major challenge when using solid-state emitters is inhomogeneous broadening, typically caused by emitter size variation and strain fields in the host material. This means that two emitters usually have non-identical emission wavelengths and thus the two QDs in our scheme cannot be easily assumed to have the same resonant transition frequencies in a realistic scenario. Here, as one of the merits of the scheme, we show the robustness of fidelity against detunings of the QDs. Such robustness has been pointed out already in Ref.~\cite{Sridharan08}, but here we provide a more detailed analysis for the case of a plasmonic nanoparticle array system. In this section, we set $\delta_{1}^{g}=\delta_{0}+\Delta\delta$ and $\delta_{2}^{g}=\delta_{0}-\Delta\delta$, and assume that $\gamma_{1}^{g}=\gamma_{2}^{g}$ and $J_{1}^{g}=J_{2}^{g}$. In figure~\ref{detunings} the fidelities are shown as the average of the detunings, $\delta_{0}=(\delta_{1}^{g}+\delta_{2}^{g})/2$, and the difference in the detunings, $\Delta\delta=(\delta_{1}^{g}-\delta_{2}^{g})/2$, are varied for $\alpha=0.5$. Here, we have taken the matching condition used in the previous section, $\alpha t_{\rm s_{1}d_{1}}^{\rm mm} +\beta t_{\rm s_{2}d_{1}}^{\rm mm} =0$. Note that this matching condition does not guarantee the optimal fidelity~\cite{footnote} but it is still sufficiently useful for observing the robustness of the fidelity against the detunings of the QDs. An optimization of the amplitude $\beta$ to reach a higher fidelity than simply using the matching condition might lead to more robustness of the fidelity against the detunings of the QDs. This, however, involves numerical calculations that make it complicated to understand the origin of the robustness and is not considered in this work. In order to understand the behaviour of the fidelities as the detunings vary, as shown in figure~\ref{detunings}~(a) and (b), it is helpful to analyse two extremal cases. The first is when the respective detunings of the two QDs have different signs while their average is zero, {\it i.e.} $\delta_{0}/g_{\rm np}=0$ and $\Delta\delta/g_{\rm np}\neq0$, labelled as line A (see figure~\ref{detunings}~(c) for level diagram), corresponding to a cut along the $y$-axis. The second is when the respective detunings of the two QDs are the same as each other, {\it i.e.} $\delta_{0}/g_{\rm np}\neq0$ and $\Delta\delta/g_{\rm np}=0$, labelled as line B (see figure~\ref{detunings}~(c) for level diagram), corresponding to a cut along the $x$-axis. The trends of the fidelity along two lines shown in figure~\ref{detunings} (d) and (e) are explained in detail in~\ref{detail2}. Here, we only discuss an interesting behaviour displayed by the fidelity around the line B.

The fidelity change in line B is such that the greater $|\delta_{0}/g_{\rm np}|$ becomes, the more robust the fidelity is against changes in $|\Delta\delta/g_{\rm np}|$. That is, the higher the overall detuning $\delta_{0}$, the better the fidelity behaves around $\Delta\delta=0$. This robustness can be understood by looking at how sensitive the various amplitudes are as they change from their values that lead to the best fidelity in line B when we vary $\Delta\delta$. Consider two examples, $\delta_{0}/g_{\rm np}=0.02$ and $\delta_{0}/g_{\rm np}=0.12$, for which the change in the amplitudes as $\Delta\omega/g_{\rm np}$ varies is shown in figure~\ref{detunings}~(f) and (g), for both $n=2$ and $n=3$. Here, it can be seen that by varying $\delta_{0}/g_{\rm np}$ (solid and dashed lines represent two different values) the resonant peaks and dips in the transmission and reflection spectrums are shifted. In addition, for a fixed value of $\Delta\delta/g_{\rm np}$ for the two chosen detunings of $\delta_{0}/g_{\rm np}$, a variation in the spectrums is induced. As we are interested in the difference in the detunings, $\Delta\omega$, one can see the values of the specturms at $\Delta\omega/g_{\rm np}=0$ are changed less sensitively by a slight variation of $\Delta\delta/g_{\rm np}$ for $\delta_{0}/g_{\rm np}=0.12$ than for $\delta_{0}/g_{\rm np}=0.02$. Such characteristics of transmission and reflection spectrums result in the robustness of the fidelity against the QD detunings $\Delta\delta/g_{\rm np}$.


\begin{figure}[t]
\centering
\includegraphics[width=11cm]{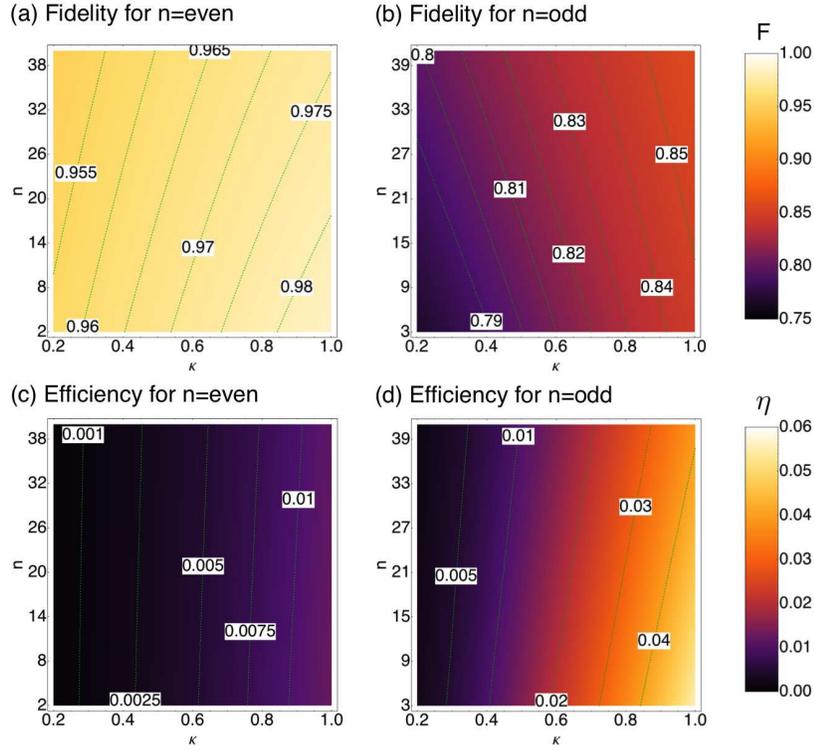}
\caption{The fidelities and efficiencies for both even and odd numbers of nanoparticles are shown as the detection efficiency (or nanowire loss), $\kappa$, varies for $\alpha=0.5$ and $g_{\rm in-out}/g_{\rm np}=0.5$. The left (right) column shows the case of $n$ even (odd), and the upper (lower) rows show the fidelity (efficiency).}
\label{detection}
\end{figure}

\subsection{Nanowire loss and detection efficiency}

As we are interested in the mapping of the input fields at the source tips to the output fields at the drain tips in the scheme, we have assumed so far that the source and drain excitations experience no loss when propagating into and out of the tip regions. However, in a realistic system, the losses in the nanowires would affect the overall performance. Here, we consider two kinds of losses in the nanowires: (i) insertion loss, which occurs when the light is transferred from a far-field source into the source nanowires and the resulting surface plasmons propagate along the nanowires up to the $1$-st or $(2n+2)$-th nanoparticle, and (ii) outcoupling loss, which occurs when surface plasmons propagate along the drain 1 nanowire and are extracted into the far-field to be detected (or equivalently the detection efficiency of an on-chip detector directly connected to $(n+1)$-th nanoparticle). For insertion loss, this is not a problem for the scheme as we can simply compensate it by increasing the intensity of the input coherent states as much as we need for the plasmon excitations at the $1$-st and $(2n+2)$-th nanoparticles. In this sense, one of the advantages of the scheme is the use of coherent states as initial `seed' states for the generation of entanglement. These enable the scheme to be stable no matter how much loss the initial coherent fields experience on input to the array system. On the other hand the second type of loss, the outcoupling loss, more significantly affects the fidelity and the efficiency. The outcoupling loss can be included as a `lumped detection efficiency', $\kappa$, by replacing ${\rm P}_{\rm d_{1}}$ with $\tilde{\rm P}_{\rm d_{1}}$, where $\tilde{\rm P}_{\rm d_{1}}=\mathbb{I} - \sum_{n=0}^{\infty}(1-\kappa)^n {\ket{n}_{\rm d_{1}}\bra{n}}$, which corresponds to the case of $n$-photons being detected by the non-photon number resolving detectors with a probability $1-(1-\kappa)^{n}$~\cite{Bell12}. Such a treatment of outcoupling loss is valid due to the fact that for an output coherent field at drain 1, say $\ket{\mu}$, the expectation value of the measurement of an ideal detection with outcoupling loss amplitude $\sqrt{\kappa}$ is equivalent to that of a lumped detection without outcoupling loss, {\it i.e.} $\bra{\mu}\tilde{\rm P}_{\rm d_{1}}\ket{\mu} = 1 - e^{-\kappa \abs{\mu}^{2}} =\bra{\sqrt{\kappa}\mu}{\rm P_{d_{1}}}\ket{\sqrt{\kappa}\mu}$. This compatibility between nanowire loss and detection efficiency allows us to treat them in the same manner. To show how much the output coupling loss (or detection efficiency) affects the entanglement generation scheme, in figure~\ref{detection} we plot the fidelity and the efficiency as the detection efficiency, $\kappa$, is varied and the number of nanoparticles, $n$, increases for $\alpha=0.5$ and $g_{\rm in-out}/g_{\rm np}=0.5$. As the detection efficiency $\kappa$ increases, the fidelity and the efficiency become generally higher. The gradients of the fidelity and the efficiency as $\kappa$ varies depend on the magnitude of the input amplitude, $\alpha$. For very small $\alpha$ (or large $\alpha$) the gradient as $\kappa$ varies is small, whereas for $\alpha$ having an intermediate value, the gradient is big. 

\subsection{Initialisation of the QDs}

In the scheme, the QDs are initially prepared in an equal superposition of $\ket{g}$ and $\ket{m}$. However, here we show that an equal superposition is not a strict requirement for the QD initialisation. When the QDs are prepared in arbitrary states, the initial state of the system is given by
\begin{equation}
\ket{\Psi_{i}^{'}} = (c_{g1} \ket{g}_{1}+c_{m1}\ket{m}_{1}) (c_{g2}\ket{g}_{2}+c_{m2}\ket{m}_{2})  \ket{\alpha}_{s_{1}} \ket{\beta}_{s_{2}}  \ket{0}_{d_{1}} \ket{0}_{d_{2}},
\end{equation}
where $\abs{c_{g1}}^{2}+\abs{c_{m1}}^{2}=1$ and $\abs{c_{g2}}^{2}+\abs{c_{m2}}^{2}=1$. 
In the ideal case of the scheme, the state $\ket{\Psi_{i}^{'}}$ is transformed into $\ket{\Psi^{'}}_{\rm QDs} = \Big\{  c_{m1} c_{m2}( \alpha t_{\rm s_{1}d_{1}}^{\rm mm} +\beta t_{\rm s_{2}d_{1}}^{\rm mm}) \ket{mm}+c_{g1}c_{m2}\beta t_{\rm s_{2}d_{1}}^{\rm gm} \ket{gm}+ c_{m1}c_{g2}\alpha t_{\rm s_{1}d_{1}}^{\rm mg} \ket{mg} \Big\}$, and by choosing $\beta$ to satisfy the matching condition $\alpha t_{\rm s_{1}d_{1}}^{\rm mm} +\beta t_{\rm s_{2}d_{1}}^{\rm mm} =0$, we have the final state
\begin{equation}
\ket{\Psi^{'}}_{\rm ent.QDs}=\frac{1}{\sqrt{N}}( c_{m1} c_{g2} \ket{mg} - c_{g1} c_{m2} \ket{gm}),
\label{Psi'}
\end{equation}
where the normalisation is given by $N=\abs{c_{m1} c_{g2}}^{2}+\abs{c_{g1} c_{m2} }^{2}$. Equation~(\ref{Psi'}) shows that $\ket{\Psi^{'}}_{\rm ent.QDs}$ is the desired ideal entangled state $\ket{\psi^-}$ as long as $ c_{m1} c_{g2} = c_{g1} c_{m2} $ is satisfied. In other words, we can achieve a high fidelity even when the two QDs are not prepared in an equal superposition. An equal superposition at the initialisation stage only guarantees a higher efficiency $\eta$, but is not required for achieving a high fidelity. This property of the scheme provides greater flexibility in the initialisation of the QDs. 


\section{Conclusion} 

In this work we have described an entanglement generation scheme between two QDs using the plasmonic modes of a metal nanoparticle array. Here, we have taken a particular nanoparticle structure consisting of two metal nanoparticle arrays which meet at a beam splitter of nanoparticles. Although Ohmic loss (energy dissipation) in metals generally induces damping in the supported plasmonic systems and limits the use of plasmonics for quantum control and state engineering, we have shown a scheme for entanglement generation that provides a robust performance against metal loss and the length of the nanoparticle arrays. Such robustness overcomes the length limit of nanoparticle arrays for use in nanoscale photonic quantum networks, thus opening up further possibilities for constructing quantum plasmonic devices on scales far below the diffraction limit. In addition, in our investigation we showed that the fidelity of the entangled QD states was robust against inhomogeneous broadenings of the QDs, implying the scheme works even when the QDs have different resonant frequencies. Through our analysis, we demonstrated that the robustness of the scheme originates from the characteristics of the transmission and reflection amplitudes of the system. The robustness against loss comes about as the fidelity mostly depends on the behaviour of the reflection coefficient of each arm and the reflection is less sensitive with nanoparticle losses since the energy never enters the array. The robustness against detunings of the QDs comes about as the operating frequency to be detected is far off-resonant with the QDs transitions. Furthermore, we discussed the effects of detection efficiency, the initialisation of the QDs, and the use of coherent states as initial states, which reveals additional versatility of the scheme in a realistic plasmonic system. While here we have concentrated on treating the main source of imperfection, that is metal loss, further works geared toward an experimental demonstration could include fabrication issues such as disorder in the resonance frequency of the nanoparticles and their coupling due to the non-ideal shape of realistic nanoparticles and the aperiodic arrangement of the array \cite{disorder}. Another fabrication issue is the error of the $50/50$ beamsplitter of nanoparticles, which should also be examined in a further study. Effects of the different life times of the ground and metastable states on generated entanglement is also an interesting future work for the use of entanglement. The techniques we have employed to describe the nanoparticle system in our work may also be helpful in further theoretical and experimental studies of plasmonic nanostructures for quantum-control applications and probing nanoscale optical phenomena. We hope that the results presented in this paper will encourage the use of metal nanoparticle structures with more complex designs for on-chip quantum networking.

\section*{Acknowledgments}

This work was supported by the Singapore's National Research Foundation and Ministry of Education, the National Research Foundation of Korea grant funded by the Korea Government (Ministry of Education, Science and Technology; grant numbers 2010-0015059 and 2010-0018295), the Leverhulme Trust, the UK's Engineering and Physical Sciences Research Council, and the European Office of Aerospace Research and Development (EOARD).

\appendix
\section{Dipole induced reflection in a metal nanoparticle array}\label{DIR}

\begin{figure}[t]
\centering
\includegraphics[width=15cm]{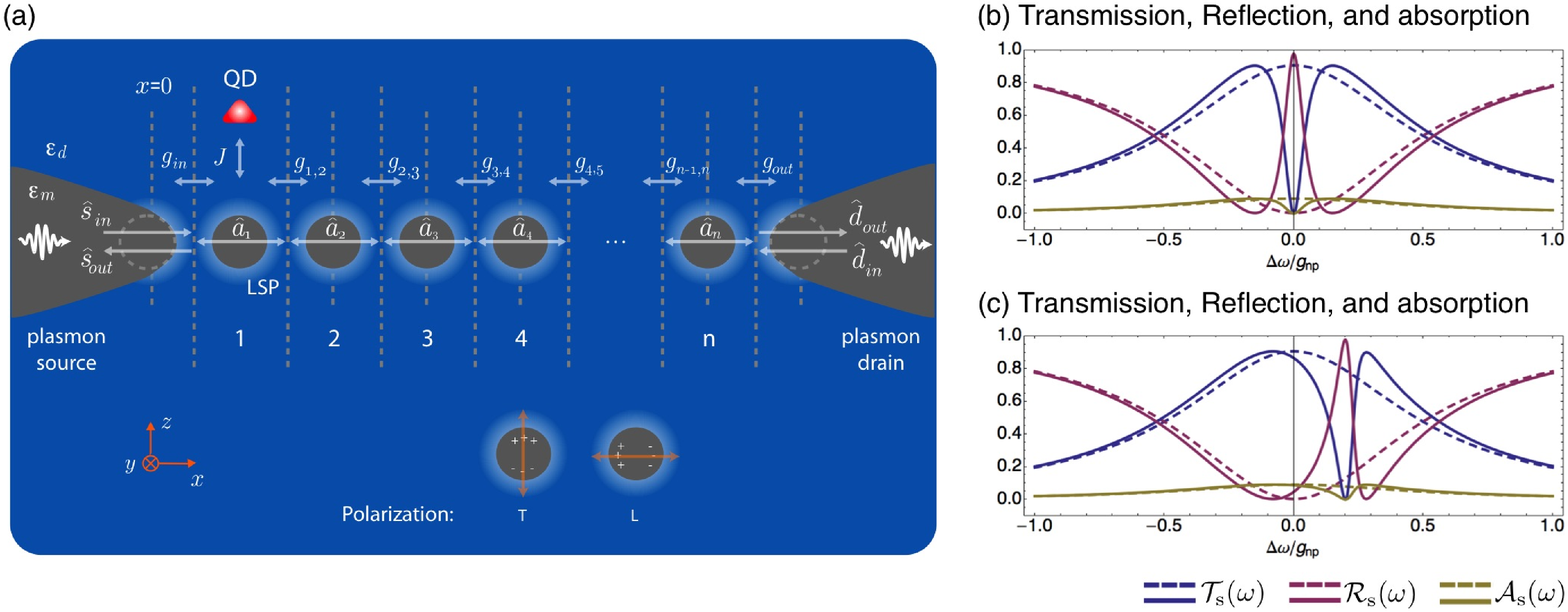}
\caption{(a) Setup for DIR. A single array of metal nanoparticles with one dipole (QD) coupled to the first nanoparticle and a tapered metal nanowire waveguide on the left-hand side (source), which focuses light to the end of its tip and excites a localised surface plasmon on the adjacent nanoparticle. The excitation propagates across the array of nanoparticles and exits via another tapered metal nanowire waveguide on the right-hand side (drain). All metal regions have permittivity $\epsilon_{m}(\omega)$ and dielectric regions have permittivity $\epsilon_{d}$, as defined in the text. Panels (b) and (c) show transmission, reflection, and nanoparticle absorption when $\delta/g_{\rm np}=0$ and $\delta/g_{\rm np}=0.2$, respectively, for $n=1$. Solid lines are for DIR (QD in $\ket{g}$), and dashed lines are for when there is no DIR (QD in $\ket{m}$). When $\Delta\omega \approx \delta$, DIR occurs, and both transmission and loss from nanoparticle dissipation are nearly zero, while reflection is close to one. 
}
\label{subsetup}
\end{figure}

Here, we provide an analysis of DIR in a single array of metal nanoparticles with one dipole (QD), which corresponds to each arm in figure~\ref{setup}~(a). The physical system is depicted in figure~\ref{subsetup}, where the $n$-th nanoparticle is connected to a drain nanowire. The Hamiltonian for the nanoparticle array is given by equation~(\ref{Hnp}), yielding the Heisenberg equations 
\begin{eqnarray}
\fl \frac{d\hat{a}_{1}}{dt} &= -(i \omega_{1} + \frac{g_{\rm s}}{2}+\frac{\Gamma_{1}}{2}) \hat{a}_{1} - i g_{1,2} \hat{a}_{2} + \sqrt{g_{\rm s}}\hat{s}_{\rm in} + \sqrt{\Gamma_{1}}\hat{b}_{\rm in,1} - i J \hat{\sigma}^{g(m)}\\
\fl \frac{d\hat{a}_{j}}{dt} &= -(i \omega_{j} +\frac{\Gamma_{j}}{2}) \hat{a}_{j} - i g_{j,j-1} \hat{a}_{j-1} - i g_{j,j+1} \hat{a}_{j+1} + \sqrt{\Gamma_{j}}\hat{b}_{\rm in,j} ~~~~~~~~ (j=2,3,\cdots,n-1)\\
\fl \frac{d\hat{a}_{n}}{dt} &= -(i \omega_{n} + \frac{g_{\rm d}}{2}+\frac{\Gamma_{n}}{2}) \hat{a}_{1} - i g_{n-1,n} \hat{a}_{n-1} + \sqrt{g_{\rm d}}\hat{d}_{\rm in} + \sqrt{\Gamma_{n}}\hat{b}_{\rm in,1}\\
\fl \frac{d\hat{\sigma}}{dt}&=-(i\Omega +\frac{\gamma}{2})\hat{\sigma}+i J\hat{a}_{1} \hat{\sigma}_{z}+\hat{f}_{\rm d},
\end{eqnarray}
where $\omega_{i}$ is the natural frequency of the field oscillation at the $i$-th nanoparticle, $g_{j,k}$ is the coupling strength between the fields of the $j$-th and $k$-th nanoparticles, $\Omega$ is the transition frequency from $\ket{g}$ to $\ket{e}$ of the dipole, $J$ is the vacuum Rabi frequency of the dipole coupled to the adjacent nanoparticle, $\Gamma_{i}$ is the damping rate of the $i$-th nanoparticle, $\gamma$ is the damping rate of the dipole, and $g_{\rm s~(d)}$ is the coupling strength of the source (drain) nanowire to its adjacent nanoparticle. The operators $\hat{a}^\dag_{i}$ ($\hat{a}_{i}$) represent the creation (annihilation) operators associated with a dipole-field excitation at $i$-th nanoparticle, $\hat{\sigma}=\ket{g}\bra{e}$, and $\hat{\sigma}_{z}=\ket{e}\bra{e}-\ket{g}\bra{g}$. The operators $\hat{s}_{\rm in~(out)}$ and $\hat{d}_{\rm in~(out)}$ represent the input (output) annihilation operators of the fields of the nanowires, respectively, which satisfy the boundary conditions $\hat{a}_{1}=\frac{1}{\sqrt{g_{\rm s}}}(\hat{s}_{\rm in}+\hat{s}_{\rm out})$ and $\hat{a}_{n}=\frac{1}{\sqrt{g_{\rm d}}}(\hat{d}_{\rm in}+\hat{d}_{\rm out})$. The input (output) field of the $i$-th nanoparticle's bath is represented by $\hat{b}_{\rm in~(out),i}$, which satisfy the boundary conditions $\hat{a}_{i}=\frac{1}{\sqrt{\Gamma_{i}}}(\hat{b}_{\rm in,i}+\hat{b}_{\rm out,i})$ for $j=1,\cdots,n$, and $\hat{f}_{\rm d}$ denotes the noise operator for the dipole.

For the above description to be valid all system parameters are restricted by the appropriate approximations mentioned in the main text of this paper. We also assume, for simplicity, that $\omega_{i}=\omega_{0}$ for $i=1,\cdots,n$, $g_{j,k}=g_{\rm np}$, $\Gamma_{i}=\Gamma_{0}$ for $i=1,\cdots,n$, and  $g_{\rm s}=g_{\rm d}=g_{\rm in-out}$. The damping rate of the dipole is assumed to be relatively small compared to other rates, so that the noise operators, $\hat{f}_{\rm d}$, can be neglected in that they do not significantly affect quantum coherence of the system. Also, here we consider either longitudinal or transverse polarisation along the array according to the direction of electron-charge-density oscillations. We impose the weak-excitation regime, where the dipole is assumed to be predominantly in the initial state $\ket{g}$. Physically, the weak-excitation limit is valid when a single-exctation pulse has a duration that is much longer than the spontaneous lifetime of the dipole~\cite{Rephaeli10}. Such an approximation has been commonly adopted in many quantum optics calculations~\cite{Waks06a, approx}. In this limit, $\hat{\sigma}_{z}$ can be replaced by its mean value $\langle \hat{\sigma}_{z} \rangle \approx -1$, so that the total coupled Heisenberg equations give rise to the scattering matrix
\begin{eqnarray} 
\hat{s}_{\rm in}^{\dagger}(\omega)&=r_{\rm s}(\omega) \hat{s}_{\rm out}^{\dagger}(\omega)+t_{\rm s}(\omega)\hat{d}_{\rm out}^{\dagger}(\omega)+\hat{f}_{\rm out, s}^{\dagger}(\omega) \\
\hat{d}_{\rm in}^{\dagger}(\omega)&=t_{\rm d}(\omega) \hat{s}_{\rm out}^{\dagger}(\omega)+r_{\rm d}(\omega)\hat{s}_{\rm out}^{\dagger}(\omega)+\hat{f}_{\rm out, d}^{\dagger}(\omega),\nonumber
\label{sm}
\end{eqnarray}
where the noise operators for the nanoparticles are $\hat{f}_{\rm out, s}^{\dagger}(\omega)=\sum_{i} t_{\rm s,b_{i}}(\omega)\hat{b}_{\rm out,i}^{\dagger}(\omega)$ and $\hat{f}_{\rm out, d}^{\dagger}(\omega)=\sum_{i} b_{\rm d,b_{i}}(\omega)\hat{b}_{\rm out,i}^{\dagger}(\omega)$. For a given input field at the source (drain) nanowires, the transmission and reflection amplitudes are represented by $t_{\rm s~(d)}$ and $r_{\rm s~(d)}$, respectively, and the coefficient $b_{\rm s~(d),b_{i} }$ denotes the transition amplitude to a bath mode at the $i$-th nanoparticle. Here, $\abs{t_{\rm s~(d)}(\omega)}^{2} + \abs{r_{\rm s~(d)}(\omega)}^{2} + \sum_{i}\abs{b_{\rm s~(d),b_{i}}(\omega)}^{2}\leqslant 1$, the left hand side does not reach unity as it is reduced by the loss of the dipole at the given frequency $\omega$ for which the dipole interacts with the nanoparticle. We define the transmission, reflection, and absorption energy as ${\cal T}_{\rm s~(d)}(\omega)=\abs{t_{\rm s~(d)}(\omega)}^{2}$, ${\cal R}_{\rm s~(d)}(\omega)=\abs{r_{\rm s~(d)}(\omega)}^{2}$, and ${\cal A}_{\rm s~(d)}(\omega)=\sum_{j}\abs{b_{\rm s~(d),b_{j}}(\omega)}^{2}$, respectively.

\subsection{Single nanoparticle}

We first consider the case of $n=1$ in order to provide a basic understanding of the physical mechanism for DIR. Arbitrary $n$ is considered in the next section. For a given input field from the source nanowire we have
\begin{eqnarray}
t_{\rm s}(\Delta\omega) &=&\frac{g_{\rm in-out} (\frac{\gamma}{2} + i(\delta-\Delta\omega))}{J^{2} + (\frac{\gamma}{2} + i(\delta-\Delta\omega))(g_{\rm in-out} +\frac{\Gamma_{0}}{2} - i\Delta\omega) } \label{ts}\\
r_{\rm s}(\Delta\omega) &=&- \frac{J^{2} + (\frac{\gamma}{2} + i(\delta-\Delta\omega))(\frac{\Gamma_{0}}{2} - i\Delta\omega) } {J^{2} + (\frac{\gamma}{2} + i(\delta-\Delta\omega))(g_{\rm in-out} +\frac{\Gamma_{0}}{2} - i\Delta\omega) } \label{rs}\\
b_{\rm s,b_{1}}(\Delta\omega) &=& \frac{ \sqrt{g_{\rm in-out}} \sqrt{\Gamma_{0}} (\frac{\gamma}{2} + i(\delta-\Delta\omega))} {J^{2} + (\frac{\gamma}{2} + i(\delta-\Delta\omega))(g_{\rm in-out} +\frac{\Gamma_{0}}{2} - i\Delta\omega) }, \label{bs}
\end{eqnarray}
where $\Delta\omega=\omega-\omega_{0}$ and $\delta=\Omega-\omega_{0}$ denote the detunings of the input field from the nanoparticle resonance and the dipole resonance from nanoparticle resonance, respectively. If there is no dipole in the proximity of the nanoparticle, {\it i.e.} $J=0$, in the ideal case ($\Gamma_{0} =0$), the field is entirely transmitted on resonance ($\Delta\omega=\omega-\omega_{0}=0$), {\it i.e.} $r_{\rm s}(\Delta\omega=0)=0$ and $t_{\rm s}(\Delta\omega=0)=1$. Once the loss in the metal nanoparticle is included, {\it i.e.} $\Gamma_{0} \neq0$, the field is no longer entirely transmitted on resonance, {\it i.e.}, $t_{\rm s}(0)\neq1$, as shown in figure~\ref{subsetup}~(b) as dashed lines. However, if there is a dipole in the proximity of the nanoparticle, {\it i.e.} $J\neq0$, then the transmission and reflection behave differently from the case of no dipole as we discuss next. 

\subsection*{Dipole and single nanoparticle in tune $(\delta=0)$} 

Consider the case where the dipole is resonant with the nanoparticle ($\delta=0$), equations~(\ref{ts})-(\ref{bs}) can be rewritten as
\begin{eqnarray}
t_{\rm s}(\Delta\omega) &=&\frac{g_{\rm in-out} (\frac{\gamma}{2} - i \Delta\omega)}{J^{2} + (\frac{\gamma}{2} - i \Delta\omega)(g_{\rm in-out} +\frac{\Gamma_{0}}{2} - i\Delta\omega) } \\
r_{\rm s}(\Delta\omega) &=&- \frac{J^{2} + (\frac{\gamma}{2} - i \Delta\omega)(\frac{\Gamma_{0}}{2} - i\Delta\omega) } {J^{2} + (\frac{\gamma}{2} - i \Delta\omega)(g_{\rm in-out} +\frac{\Gamma_{0}}{2} - i\Delta\omega) } \\
b_{\rm s,b_{1}}(\Delta\omega) &=& \frac{ \sqrt{g_{\rm in-out}} \sqrt{\Gamma_{0}} (\frac{\gamma}{2} - i \Delta\omega)} {J^{2} + (\frac{\gamma}{2} - i \Delta\omega)(g_{\rm in-out} +\frac{\Gamma_{0}}{2} - i\Delta\omega) }.
\end{eqnarray}
In this case, when $2J^{2}/\gamma \gg g_{\rm in-out}+\Gamma_{0}/2$ we have $t_{\rm s}(0)\approx0$ and $r_{\rm s}(0) \approx-1$, with a $\pi$ phase shift on resonance ($\Delta\omega=0$), so that the field is totally reflected, similar to the frequency selective perfect mirror described by Shen and Fan~\cite{Shen05}. This is a quantum interference effect called {\it dipole induced reflection} (DIR). It is based on the same mechanism as {\it dipole induced transparency} (DIT) originally introduced by Waks~{\it et al.}~\cite{Waks06a}, in which the cavity-dipole system is driven by an external field and the cavity field destructively interferes with the excited state population of the dipole. Here, a metal nanoparticle serves as a bad cavity, where cavity decay rate $\Gamma_{0}$ is larger than the dipole decay rate $\gamma$. Nevertheless, in this bad cavity regime, dipole induced reflection (or transmission) still provides strong dispersive properties (as shown in figure~\ref{subsetup}~(b) as solid lines), which is one of the interesting merits of the original DIT scheme, as pointed out by Waks {\it et al.}~\cite{Waks06a}. Our plasmonic system for $n=1$ is mathematically equivalent to driving a double sided cavity with an incident field.

In order to understand the origin of the dispersive properties, we use the Purcell factor, defined as $F_{p}=\frac{2J^{2}}{\gamma}\frac{1}{g_{\rm in-out}+\Gamma_{0}/2}$ (proportional to the ratio of the dipole decay rate into the nanoparticle, $J$, to the bare dipole decay rate, $\gamma$). In terms of the Purcell factor, the transition amplitudes can be rewritten on resonance as $t_{\rm s}(\Delta\omega=0) = t_{0}/(F_{p}+1)$, $r_{\rm s}(\Delta\omega=0) = -(F_{p}-r_{0})/(F_{p}+1)$ and $b_{\rm s,b_{1}}(\Delta\omega=0)=a_{0}/(F_{p}+1)$, where $t_{0}=\frac{g_{\rm in-out}}{g_{\rm in-out}+\Gamma_{0}/2}$, $r_{0}=-\frac{\Gamma_{0}/2}{g_{\rm in-out}+\Gamma_{0}/2}$, and $a_{0}=\frac{\sqrt{g_{\rm in-out}}\sqrt{ \Gamma_{0}}}{g_{\rm in-out}+\Gamma_{0}/2}$ are the transmission, reflection, and absorption amplitudes for a bare nanoparticle in the absence of the dipole. One can see that in order to have DIR a large Purcell factor is required, {\it i.e.} $F_{p} \gg1$. However, as in DIT, we do not need the full normal mode splitting condition $J > g_{\rm in-out} +\Gamma_{0}/2$, known as the high-Q cavity regime. We can achieve a large Purcell factor when $\gamma \ll  g_{\rm in-out} +\Gamma_{0}/2$ even for much smaller values of $J$. This is known as the low-Q cavity regime, where the coupling strength $J$ between the nanoparticle and the dipole is less than the nanoparticle decay rate $\Gamma_{0}$. 

When the nanoparticle plasmonic excitations and the $\ket{g}$-$\ket{e}$ transition of the dipole are in tune ($\delta=0$), the overall spectrum of the reflection amplitude is always symmetric with respect to $\Delta\omega=0$, at which the local maximum is located, as shown in figure \ref{subsetup}~(b). Here, the dipole dissipation rate has a stronger effect on the reflection (transmission) of an on-resonance excitation than the nanoparticle dissipation rate does. When $\gamma=0$, a plasmon is still completely reflected, even with the presence of the nanoparticle loss $\Gamma_{0}$, whereas when $\gamma\neq0$, the dissipation of the dipole interrupts the quantum interference, so that the nanoparticle is marginally excited for a plasmon at $\omega=\Omega$ and the loss via the nanoparticle increases. Two minima (maxima) in the reflection (transmission) spectrum are located at $\Delta\omega \approx \pm J$, in the limit of a large Purcell factor $F_{p}\gg1$, which corresponds to the Rabi-split frequencies $\omega= \Omega \pm J$. We find that the linewidth of the broadest transmission peak is the nanoparticle linewidth whereas the linewidth of the narrow dip corresponds to the linewidth of the dipole dressed by the nanoparticle excitation. This quantum interference that produces the narrow reflection window with a simultaneously strong dispersion results in a significant enhancement of the group delay and the possibility of a slowdown or `storage' of light, as in electromagnetic induced transparency~\cite{EIT}.

\subsection*{Dipole and single nanoparticle detuned $(\delta\neq0)$} 

Consider the case where the dipole is not resonant with the nanoparticle, {\it i.e.} $\delta\neq0$. When the dipole transition frequency $\Omega$ is detuned slightly away from the nanoparticle frequency $\omega_{0}$, the spectrum of the reflection and transmission amplitudes become asymmetric, which is known as a Fano resonance, as discussed by Shen and Fan~\cite{Shen05}. The local maximum (minimum) in the reflection (transmission) spectrum is located at $\Delta\omega=\delta$ in the limit of a large Purcell factor $F_{p}\gg1$, regardless of the detuning between the dipole and the nanoparticle, as shown by the solid lines of figure \ref{subsetup}~(b) and (c). On the other hand, when the dipole is very far detuned from the nanoparticle resonance frequency, the dipole is essentially decoupled from the nanoparticle field, so that the reflection (transmission) properties are determined by the nanoparticle only, and the reflection (transmission) spectrum dips down to zero at the nanoparticle frequency $\omega=\omega_{0}$, as shown by the dashed lines in figure~\ref{subsetup}~(b) and (c). This feature could be exploited to achieve a fast single-excitation switch: for an incoming photon with frequency $\omega=\omega_{0}$, the transmission is $1$ when the dipole is in tune with the nanoparticle ($\Omega=\omega_{0}$), while the transmission is essentially $0$ when the dipole is far-detuned. Thus by tuning the transition frequency of the dipole, the single-plasmon transport can be regulated and the setup acts as a single-plasmon (or photon in the far field) switch, as pointed out in Ref.~\cite{Waks06a}.


\begin{figure}[t]
\centering
\includegraphics[width=13cm]{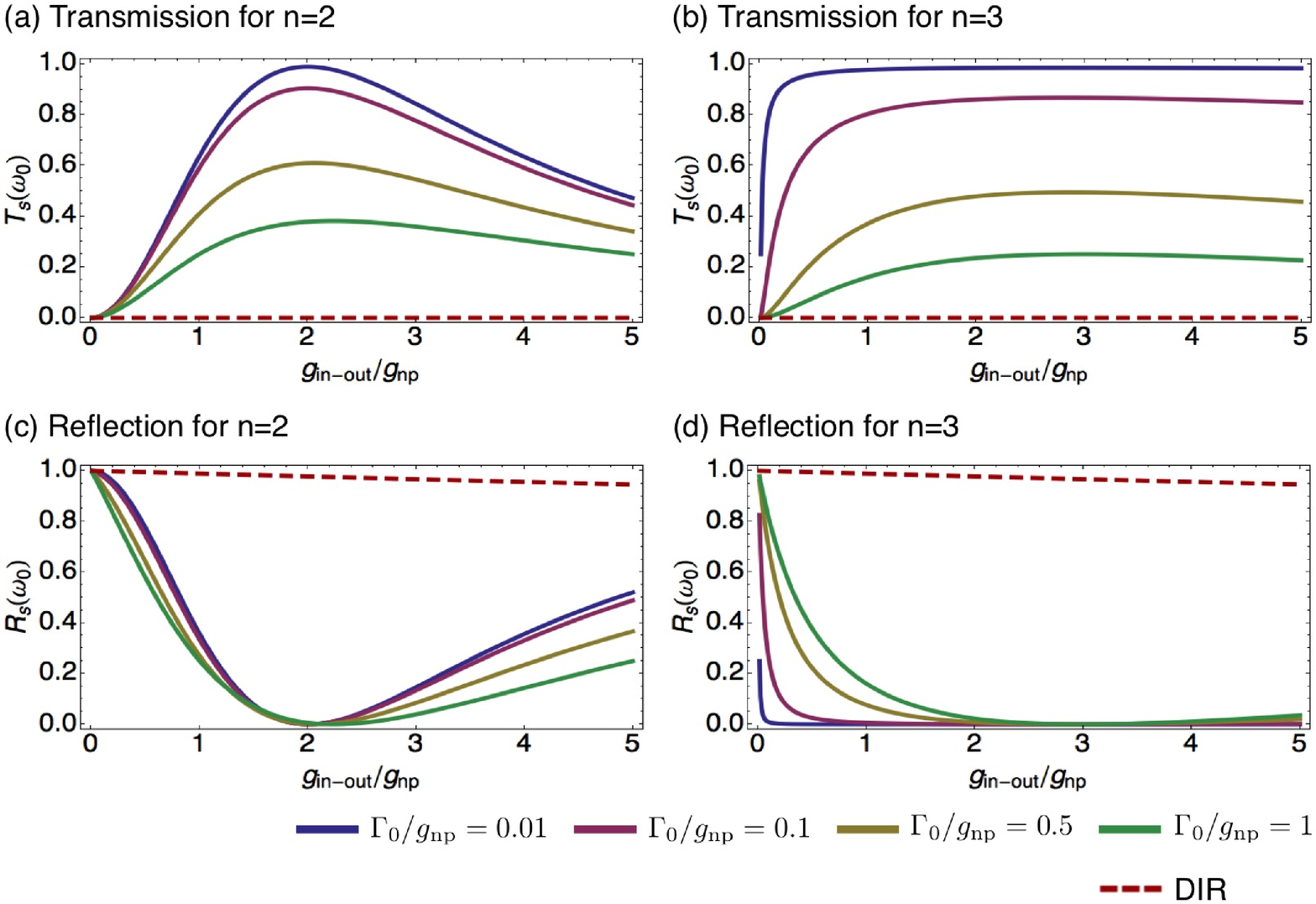}
\caption{DIR for an arbitrary number of nanoparticles. The transmission ${\cal T}_{\rm s}(\omega)$ (upper) and reflection ${\cal R}_{\rm s}(\omega)$ (lower) at $\Delta\omega=0$ is shown as $g_{\rm in-out}/g_{\rm np}$ is varied for $n=2$ (left) and $n=3$ (right), as representatives of even and odd number of nanoparticles respectively. If the dipole is decoupled from its adjacent nanoparticle (solid lines), their behaviour as $g_{\rm in-out}/g_{\rm np}$ is varied becomes different; the transmission (reflection) has a maximum (minimum) value at $g_{\rm in-out}/g_{\rm np}=2$ for $n=2$, whereas the transmission (reflection) rises (drops) quickly as $g_{\rm in-out}/g_{\rm np}$ is increased for $n=3$. Furthermore the transmission and reflection are quite sensitive to $\Gamma_{0}/g_{\rm np}$ changing. On the other hand, if the dipole with a detuning of $\delta/g_{\rm np}=0$ is coupled to its adjacent nanoparticle (dashed lines), DIR can be observed when $F_{p} \gg 1$ for both $n=2$ and $n=3$. Furthermore, the transmission and reflection are no longer sensitive to $\Gamma_{0}/g_{\rm np}$ changing; the transmission is nearly zero regardless of $g_{\rm in-out}/g_{\rm np}$, and the reflection only slightly drops from 1 as $g_{\rm in-out}/g_{\rm np}$ increases due to the decay of the dipole, $\gamma$. As the number of nanoparticles is increased further, similar trends to those of increasing the metal loss are seen, since the effects of increasing $n$ are equivalent to increasing $\Gamma_{0}/g_{\rm np}$ for a fixed number of nanoparticles, as discussed in the main text.}
\label{TRginout}
\end{figure}

\begin{figure}[t]
\centering
\includegraphics[width=14cm]{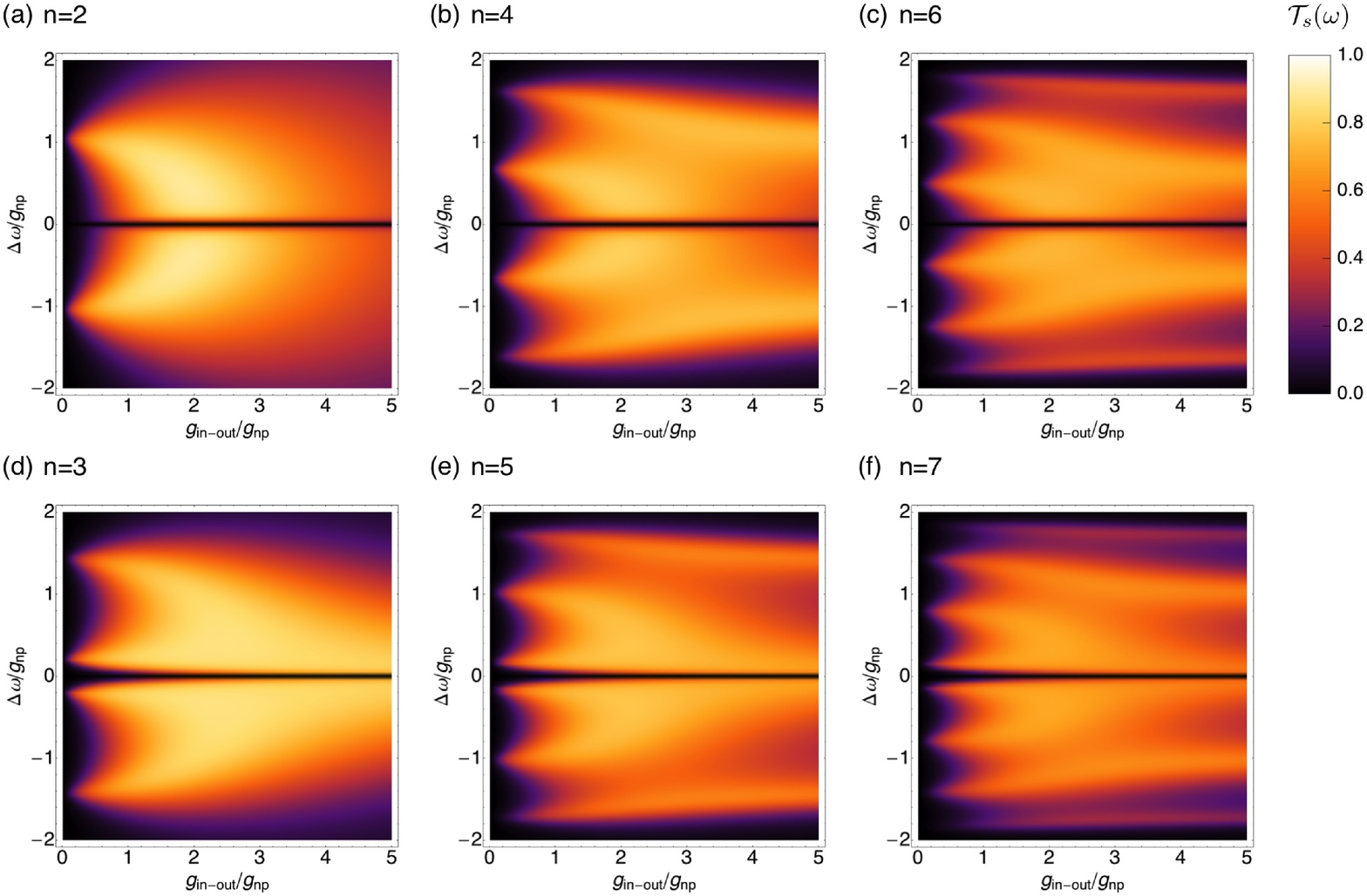}
\caption{The transmission, ${\cal T}_{\rm s}(\omega)$, as the nanoparticle detuning frequency $\Delta\omega/g_{\rm np}$ and input-output coupling $g_{\rm in-out}/g_{\rm np}$ are varied for arrays of $n=2,~3,~4,~5,~6$, and $7$ nanoparticles. Regardless of the parity of the nanoparticle number, the early increase of $g_{\rm in-out}/g_{\rm np}$ enables the off-resonant transfer from the source nanowire to the first nanoparticle, whereas its late increase leads to strong coupling as if the first nanoparticle becomes the extended `tip' of the nanowire. Thus the large $g_{\rm in-out}/g_{\rm np}$ implies that the number of nanoparticles is effectively reduced to $n-2$. Here, DIR occurs at $\Delta\omega=0$, when the dipole detuning is given by $\delta/g_{\rm np}=0$, and the transmission becomes nearly zero regardless of $g_{\rm in-out}/g_{\rm np}$ and the number of nanoparticles.}
\label{Tginoutomega}
\end{figure}

\subsection{Arbitrary number of nanoparticles}

We now increase the number of nanoparticles in the array, while keeping the dipole coupled to the first nanoparticle, as shown in figure~\ref{subsetup}. As in the case of $n=1$, we have transmission, reflection, and absorption amplitudes from the scattering matrix given in equation~(\ref{sm}). If the dipole is decoupled from nanoparticle, the transmission and the reflection amplitudes follow the characteristic of a metal nanoparticle array studied in Ref~\cite{Lee12}, which have different dependencies on $g_{\rm in-out}$. We classify their behaviour into two groups: even and odd numbers of nanoparticles in the array. As a representative of even and odd numbers of nanoparticles, we consider the case of $n=2$ and $n=3$. As shown in~\cite{Lee12}, for an odd number of nanoparticles, there is always a resonance at the natural frequency $\omega_{0}$, whereas for an even number of nanoparticles, a resonance property at the natural frequency $\omega_{0}$ depends on the magnitude of $g_{\rm in-out}/g_{\rm np}$. Thus, the transmission properties as $g_{\rm in-out}/g_{\rm np}$ is varied at $\Delta\omega=0$ are different for $n=2$ and $n=3$ when the dipole is decoupled from nanoparticle, as shown in figure~\ref{TRginout}~(a) and (b). Here, the transmission (reflection) has a maximum (minimum) value at $g_{\rm in-out}/g_{\rm np}=2$ for $n=2$, whereas the transmission (reflection) rises (drops) quickly as $g_{\rm in-out}/g_{\rm np}$ is increased for $n=3$. In addition, in figure~\ref{TRginout} we show the dependence of the transmission and reflection with varying amount of metal loss for $n=2$ and $n=3$. The dependence on the metal loss is more sensitive for the transmission than for the reflection, regardless of the parity of the number of nanoparticles, which is the main reason for the robustness against the metal loss mentioned in the main text. As the metal loss is increased, the transmissions become lower; for $n=2$ (or any even number) they still have a maximum near $g_{\rm in-out}/g_{\rm np}\approx 2$, which is reflected in the efficiency as shown in figure~\ref{fidelity loss} (c) of the main text. On the other hand, for $n=3$ (or any odd number) the transmissions are still flat as $g_{\rm in-out}/g_{\rm np}$ is increased, which is also reflected in the efficiency, as shown in figure \ref{fidelity loss} (d) of the main text. The reflection amplitudes shown in ~\ref{TRginout}~(c) and (d) are highly related to the transition amplitude $t_{\rm s_{1}s_{1}}^{gm}$ shown in figure~\ref{coefficient loss}~(a) and (b), enabling the scheme to be robust to the metal loss. 

On the other hand, if the nanoparticle is coupled to the dipole whose detuning is zero, $\delta/g_{\rm np}=0$, DIR can be observed at $\Delta\omega=0$ for $n=2$ and $n=3$ when $F_{p} \gg 1$. In figure~\ref{TRginout}, while the transmission is nearly zero (dashed lines in (a) and (b)), the reflection (dashed lines in (c) and (d)) is nearly one and decreased only slightly with $g_{\rm in-out}/g_{\rm np}$, which is related to the coefficient $t_{\rm s_{1}s_{1}}^{mg}$ in figure~\ref{coefficient loss}~(a) and (b). In addition, when DIR occurs, both the transmission and reflection amplitudes are largely insensitive to nanoparticle losses, as expected from the case of $n=1$. In figure~\ref{Tginoutomega}, the spectral profile of the transmission is presented as $\Delta\omega/g_{\rm np}$ and $g_{\rm in-out}/g_{\rm np}$ are varied for $n=2,~3,~4,~5,~6,$ and $7$, in the presence of metal loss $\Gamma_{0}/g_{\rm np}=0.1$, as in Ref.~\cite{Lee12}. Note that on resonance ($\Delta \omega=0$) DIR can be observed regardless of $g_{\rm in-out}/g_{\rm np}$ and the number of nanoparticles, where the transmission becomes nearly zero for $\Delta\omega/g_{\rm np}=0$, with the dipole detuning given as $\delta/g_{\rm np}=0$.

\section{Pulse width of input coherent fields}\label{pulse}

Here, we investigate the effects of the pulse width of the coherent field injected into the source nanowires. We start by focusing on the coherent field that comes from the source nanowire on the left-hand side, whose multi-mode coherent state can be described by $\ket{\{\alpha\}}_{\rm s_{1}} = \mathrm{exp}(\hat{s}_{{\rm in}, \alpha}^{\dagger}-\hat{s}_{{\rm in}, \alpha}) \ket{0}_{\rm s_{1}}$, where the wavepacket operators are $\hat{s}_{{\rm in}, \alpha}^{\dagger} = \int_{-\infty}^{\infty} {\rm d}\omega \alpha(\omega)\hat{s}^\dag_{\rm in}(\omega)$, with $\int_{-\infty}^{\infty} {\rm d}\omega |\alpha(\omega)|^{2}=\langle \hat{n}_{\alpha} \rangle$~\cite{Loudon00}. For concreteness, we consider a Gaussian wave packet with spectral amplitude profile $\alpha(\omega)=(2\pi \sigma_{\alpha}^{2})^{-1/4} e^{-(\omega_{0}-\omega)^{2}/4\sigma_{\alpha}^{2}}$, where $\omega_{0}$ is the central frequency and $\sigma_{\alpha}=\delta\omega_{\alpha}/(2\sqrt{2 {\rm ln} 2})$ is the standard deviation corresponding to a full width at half maximum bandwidth $\delta\omega_{\alpha}$ for the spectral intensity profile $\abs{\alpha(\omega)}^{2}$. 

Such a coherent field pulse should be carefully employed with appropriate constraints for the analysis of this work to be valid. First, for the monochromatic approximation ($\Delta\omega=0$) to be valid, $\delta\omega_{\alpha}$ should be narrow enough such that as the entanglement generation scheme is carried out, each amplitude of the system hardly changes (or varies slowly). This is equivalent to non-dispersive transfer of the plasmons in the metal nanoparticle array~\cite{Lee12}. Second, the higher-order photon number contributions that induce decoherence impose a constraint of $\langle \hat{n}_{\alpha} \rangle \ll1$ for a high-fidelity entangled state to be generated between the QDs. Third, the analysis of the protocol using DIR is valid in the weak excitation regime where the QDs are unsaturated, {\it i.e.} $\langle \hat{\sigma}_{z}^{g}(t) \rangle \approx -1$ which is equivalent to $\langle \hat{\sigma}_{1}^{g\dagger} \hat{\sigma}_{1}^{g} \rangle \ll 1$. If the amplitude of the input coherent field is large enough such that the $g$-$e$ transitions of the QDs are saturated, then they will lead to an optical nonlinearity and linewidth broadening~\cite{Englund05}, and equation~(\ref{sigma1}) can no longer be treated in the linear regime. To investigate the implications of this weak excitation limit, we use the Heisenberg equations of motion for the first nanoparticle and QD$_{1}$, and ignore the field operator for the second nanoparticle since it never has the chance to be excited when DIR occurs. Thus, equation~(\ref{a1}) and (\ref{sigma1}) can be rewritten in the weak excitation regime as
\begin{eqnarray}
\frac{d\hat{a}_{1}}{dt} &= -(i \omega_{0} + \frac{g_{\rm in-out}}{2}+\frac{\Gamma_{0}}{2}) \hat{a}_{1} + \sqrt{g_{\rm in-out}}\hat{s}_{\rm in,1} + \sqrt{\Gamma_{1}}\hat{b}_{\rm in,1} - i J_{1}^{g} \hat{\sigma}_{1}^{g}\\
\frac{d\hat{\sigma}_{1}^{g}}{dt}&=-(i\Omega_{1}^{g} +\frac{\gamma_{1}^{g}}{2})\hat{\sigma}_{1}^{g}-i J_{1}^{g}\hat{a}_{1}.
\end{eqnarray}
Eliminating $\hat{a}_{1}$ from the above equations when $(\frac{g_{\rm in-out}}{2}+\frac{\Gamma_{0}}{2})\frac{\gamma^{g}}{2} \ll {J_{1}^{g}}^{2}$, we have at resonance ($\Delta\omega \approx 0$),
\begin{equation}
\Bigg(  -i\delta_{1}^{g}(\frac{g_{\rm in-out}}{2}+\frac{\Gamma_{0}}{2}) -{J_{1}^{g}}^{2} \Bigg) \hat{\sigma}_{1}^{g} = - i J_{1}^{g} \sqrt{g_{\rm in-out}}\hat{s}_{\rm in,1} - i J_{1}^{g} \sqrt{\Gamma_{0}}\hat{b}_{\rm in,1}.
\end{equation}
The above equation can be multiplied by its conjugate to give
\begin{equation}
\langle \hat{\sigma}_{1}^{g\dagger}\hat{\sigma}_{1}^{g}\rangle \approx 
\frac{{J_{1}^{g}}^{2} g_{\rm in-out}}
{ {J_{1}^{g}}^{4}+\delta_{1}^{2} \Big(\frac{g_{\rm in-out}}{2}+\frac{\Gamma_{0}}{2} \Big)^{2}}
\langle \hat{s}_{\rm in,1}^{\dagger}\hat{s}^{\phantom{\dag}}_{\rm in,1}\rangle,
\end{equation}
where $\langle \hat{b}_{\rm in,1}^{\dagger}\hat{b}^{\phantom{\dag}}_{\rm in,1}\rangle \approx0$, $\langle \hat{s}_{\rm in,1}^{\dagger}\hat{b}^{\phantom{\dag}}_{\rm in,1}\rangle \approx0$, and $\langle \hat{b}_{\rm in,1}^{\dagger}\hat{s}^{\phantom{\dag}}_{\rm in,1}\rangle \approx0$ are assumed, and $\langle \hat{\sigma}_{1}^{g\dagger}\hat{\sigma}_{1}^{g}\rangle$ represents the probability of QD$_{1}$ being in the excited state. 
Here, $\langle \hat{s}_{\rm in,1}^{\dagger}\hat{s}^{\phantom{\dag}}_{\rm in,1}\rangle$ is identified as the total flux of photons in the input field $\ket{\alpha}_{\rm s_{1}}$ of frequency $\omega_{0}$. In the weak excitation limit, $\langle \hat{\sigma}_{1}^{g\dagger}\hat{\sigma}_{1}^{g}\rangle \ll 1$, which finally puts a limit on the excitation number $\langle \hat{n}_{\alpha} \rangle$ of
\begin{equation}
\frac{{J_{1}^{g}}^{2}}{g_{\rm in-out}}+\frac{\delta_{1}^{2}(g_{\rm in-out} + \Gamma_{0})^{2}}{4{J_{1}^{g}}^{2}g_{\rm in-out}} \gg \frac{\abs{\alpha}^{2}}{\delta\tau_{\alpha}} \sim \frac{\langle \hat{n}_{\alpha} \rangle}{\delta\tau_{\alpha}},
\label{WEL}
\end{equation}
where $\delta\tau_{\alpha}$ is the pulse width of $\ket{\{\alpha\}}_{\rm s_{1}}$. Note that this third constraint of the weak field excitation approximation can also cover the first and second constraints of sufficiently narrow bandwidth $\delta\omega_{\alpha}$ (or equivalently long pulse width $\delta\tau_{\alpha}$) and low excitation number $\langle \hat{n}_{\alpha} \rangle \ll 1$. Similar constraints on $\langle \hat{n}_{\beta} \rangle$ for $\ket{\{ \beta \}}_{\rm s_{2}}$ can be derived from the Heisenberg equations of motion for the $(n+2)$-th nanoparticle and QD$_{2}$.


\section{Details for robust fidelity against metal loss}\label{detail}
Here, we explain the behaviours of fidelity and efficiency of section 4.1 in detail by analysing the final state $\ket{\Psi_{f}}$ given in equation~(\ref{final state}). First, let us consider the fidelity. For the case of resonant QDs ($\delta_{1}^{g}=\delta_{2}^{g}=0$), we have the coherent amplitudes for drain 1 as $\mu_{1}^{gm}=-\mu_{1}^{mg}$, a condition required to generate a perfect singlet state, $\ket{\psi^{-}}$, as seen in the discussion of the `limiting scenario' in section 3. We also have the coherent amplitudes for drain 2 as $\mu_{2}^{gm}=\mu_{2}^{mg}$, so that the corresponding term in equation~(\ref{final state}) can be factored out from $\ket{\Psi_{f}}$. On the other hand, the coherent amplitudes for the other output modes, corresponding to the sources, $\xi_{1}^{gm}~(=-i\xi_{2}^{mg})$ and $\xi_{1}^{mg}~(=-i\xi_{2}^{gm})$, are not equal to each other and cannot be factored out, which leads to dephasing of the QD state. In addition, all of the bath modes (modelling loss at the nanoparticles) are also involved in dephasing the QD state in a similar way. These bath modes dominate the value of the fidelity for large values of $\Gamma_{0}/g_{\rm np}$ as $\rho_{23}~(=\rho_{32}^{*})\rightarrow 0$. Nevertheless, the dephasing effects of the bath modes and their impact on the fidelity are relatively weak compared to the source modes until large losses are incurred in the system, {\it i.e.} for $\Gamma_{0}/g_{\rm np} > 1$. For these reasons, both the coherent amplitudes for the source output modes, $\xi_{1}^{gm}$ and $\xi_{1}^{mg}$, are predominantly responsible for the behaviour of the fidelity before $\Gamma_{0}/g_{\rm np}$ becomes very large. Thus, we focus on investigating the behaviour of the amplitudes $\xi_{1}^{gm}$ and $\xi_{1}^{mg}$. Due to the beam splitter, the transition amplitudes $t_{\rm s_{2}s_{1}}^{gm}=0$ and $t_{\rm s_{2}s_{1}}^{mg}=0$, so that $\xi_{1}^{gm}=\alpha t_{\rm s_{1}s_{1}}^{gm}$ and $\xi_{1}^{mg}=\alpha t_{\rm s_{1}s_{1}}^{mg}$, where $t_{\rm s_{1}s_{1}}^{gm}$ is the reflection coefficient when DIR occurs for QD$_1$, while $t_{\rm s_{1}s_{1}}^{mg}$ is related to the reflection coefficient of a single arm in the absence of QD$_1$. Thus, the more similar $t_{\rm s_{1}s_{1}}^{mg}$ and $t_{\rm s_{1}s_{1}}^{gm}$ are to each other, the more similar $\xi_{1}^{gm}$ and $\xi_{1}^{mg}$ are to each other, and the output states in the corresponding modes can be factored out from the state $\ket{\Psi_{f}}$ in equation~(\ref{final state}). This leads to a high fidelity. Otherwise, the more $t_{\rm s_{1}s_{1}}^{mg}$ and $t_{\rm s_{1}s_{1}}^{gm}$ are dissimilar to each other, the more they cause dephasing. 

\begin{figure}[t]
\centering
\includegraphics[width=13cm]{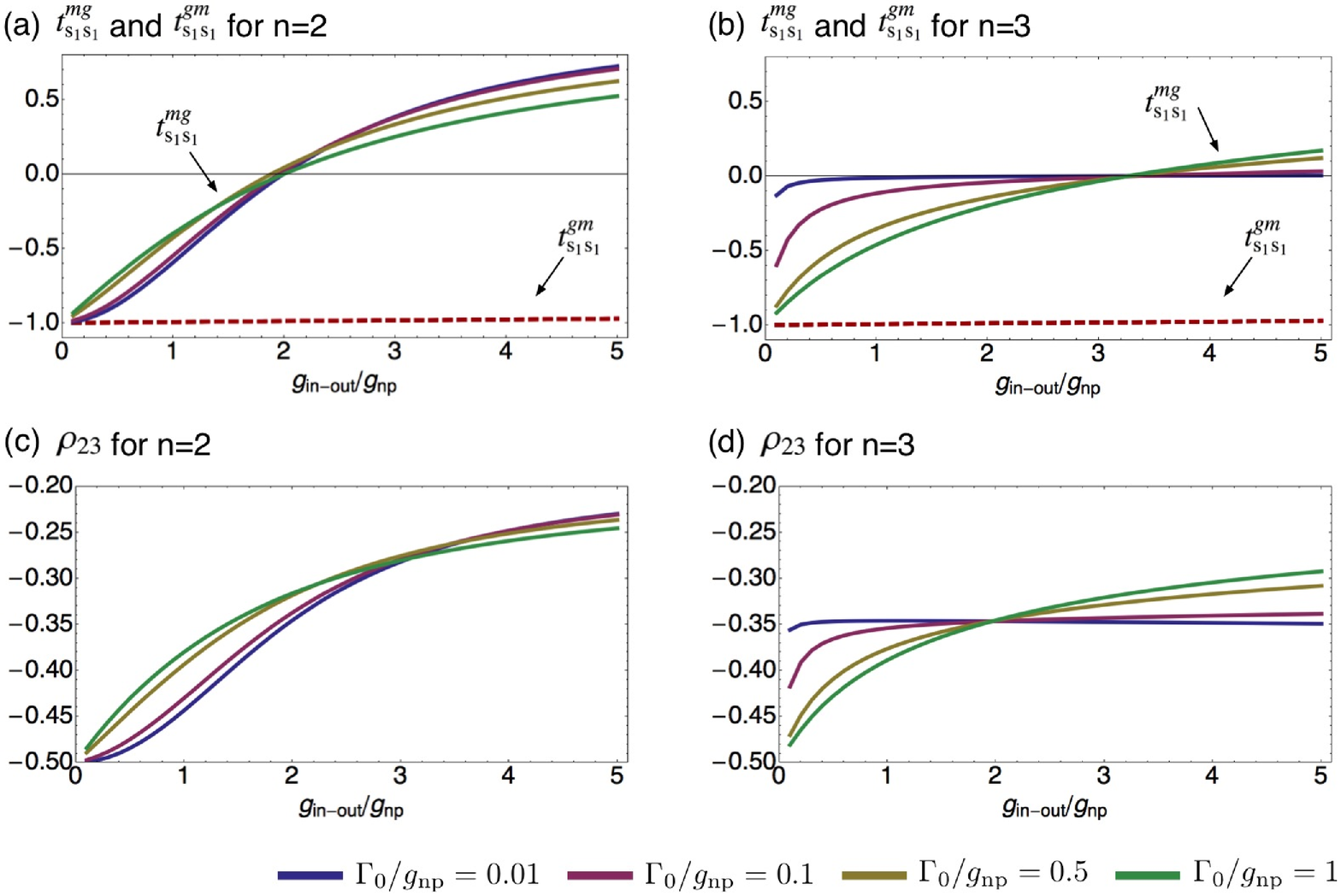}
\caption{Origin of the robustness of the fidelity. In (a) and (b) the transition amplitudes $t_{\rm s_1s_1}^{mg}$ and $t_{\rm s_1s_1}^{gm}$ are shown. In (c) and (d) the density matrix entry $\rho_{23}$ is shown. In all plots $g_{\rm in-out}/g_{\rm np}$ is varied for set amounts of metal loss: $\Gamma_{0}/g_{\rm np}=0.01, 0.1, 0.5$, and $1$.}
\label{coefficient loss}
\end{figure}

To compare $\xi_{1}^{gm}$ and $\xi_{1}^{mg}$, we choose as an example $\alpha=0.5$ in figure~\ref{coefficient loss}, and show the dependence of $t_{\rm s_{1}s_{1}}^{gm}$ and $t_{\rm s_{1}s_{1}}^{mg}$ on the input/output coupling $g_{\rm in-out}/g_{\rm np}$ for a set of values of metal loss. In figure~\ref{coefficient loss}~(a) and (b), the coefficient $t_{\rm s_{1}s_{1}}^{gm}$ (dashed line), which is related to the occurrence of DIR, does not change appreciably with respect to the increase in loss. This is because the first nanoparticle is not excited due to quantum interference when DIR occurs (see~\ref{DIR}.2), similar to a waveguide coupled to a cavity with a dipole~\cite{Sridharan08}. On the other hand, the coefficient related to the absence of a QD, $t_{\rm s_{1}s_{1}}^{mg}$, exhibits the characteristics of a nanoparticle array, where the dependence on $g_{\rm in-out}/g_{\rm np}$ is different for even and odd numbers of nanoparticles, as seen in~\ref{DIR}.2. For $n=2$, the coefficient $t_{\rm s_{1}s_{1}}^{mg}$ diverges from $t_{\rm s_{1}s_{1}}^{gm}$ as $g_{\rm in-out}/g_{\rm np}$ increases and reveals a slight variation in $\Gamma_{0}/g_{\rm np}$, which explains the falling fidelity as $g_{\rm in-out}/g_{\rm np}$ increases and robustness of the fidelity with increasing $\Gamma_{0}/g_{\rm np}$, as shown in figure \ref{fidelity loss} (a). For $n=3$, the coefficient $t_{\rm s_{1}s_{1}}^{mg}$ also diverges from $t_{\rm s_{1}s_{1}}^{gm}$ as $g_{\rm in-out}/g_{\rm np}$ increases, which again explains the fidelity decay as $g_{\rm in-out}/g_{\rm np}$ increases in figure~\ref{fidelity loss}~(b). However, contrary to the case of $n=2$, the coefficient $t_{\rm s_{1}s_{1}}^{mg}$ moves closer to $t_{\rm s_{1}s_{1}}^{gm}$ as $\Gamma_{0}/g_{\rm np}$ increases when $g_{\rm in-out}/g_{\rm np}$ is small, and shows a bigger variation compared to the $n=2$ case when $g_{\rm in-out}/g_{\rm np}$ is increased, as shown in figure~\ref{coefficient loss}~(b). This analysis explains the slight increase in fidelity with $\Gamma_{0}/g_{\rm np}$ for small $g_{\rm in-out}/g_{\rm np}$, and the robust fidelity with $\Gamma_{0}/g_{\rm np}$ for large $g_{\rm in-out}/g_{\rm np}$ that is shown in figure~\ref{fidelity loss}~(b). While $\xi_{1}^{gm}$ and $\xi_{1}^{mg}$ are approximately responsible for the fidelity when $\delta_{1}^{g}=\delta_{2}^{g}=0$, the entry $\rho_{23}$ in the QD density matrix is the only entry that directly determines the fidelity among the different entries of $\rho_{\rm QDs}$, since $\rho_{22}=\rho_{33}=1/2$ when $\delta_{1}^{g}=\delta_{2}^{g}=0$. In figure~\ref{coefficient loss}~(c) and (d), we plot $\rho_{23}$. By comparing the behaviour of $\rho_{23}$ in these plots with that of the fidelity in figure~\ref{fidelity loss}~(a) and (b), it can be seen that as the real part of $\rho_{23}$ goes to $-\frac{1}{2}$, the fidelity increases. Thus, the behaviour of $\rho_{23}$ follows the trend of $t_{\rm s_{1}s_{1}}^{mg}$ with increasing $\Gamma_{0}/g_{\rm np}$ and $g_{\rm in-out}/g_{\rm np}$, except for the intermediate regime of $g_{\rm in-out}/g_{\rm np}$. Here, there is an interplay of other field modes in the system, which become significant and make a small difference in the trend of $\rho_{23}$ compared to that of the source coherent amplitudes $\xi_{1}^{gm}$ and $\xi_{1}^{mg}$.

Secondly, let us consider the efficiency. In contrast to the fidelity, the efficiency depends only on the amplitudes $\mu_{1}^{gg}$, $\mu_{1}^{mm}$, $\mu_{1}^{gm}$, and $\mu_{1}^{mg}$, as seen in equation (\ref{efficiency}), which are mostly related to the transmission amplitudes of the two arms. In appendix~\ref{DIR}.2, the square of the transmission amplitudes are presented in figure~\ref{TRginout}, where for $n=2$ the transmission is found to have its highest value at $g_{\rm in-out}/g_{\rm np}=2$, whereas for $n=3$ the transmission quickly increases with increasing $g_{\rm in-out}/g_{\rm np}$. Such characteristics of the nanoparticle array are reflected in the efficiencies for both $n=2$ and $n=3$, as shown in figure~\ref{fidelity loss}~(c) and (d). Higher values of the transmission as $g_{\rm in-out}/g_{\rm np}$ is varied enable the probability of the injected plasmons to exit via drain nanowires to become more likely, so that the efficiency to detect excitations at drain 1 increases (compare figure~\ref{fidelity loss}~(c) and (d) with figure~\ref{TRginout}~(a) and (b)). This shows that the efficiency does not have simple trade-off with the fidelity in figure~\ref{fidelity loss}.

\section{Details for robust fidelity against detunings of the QDs}\label{detail2}
In line A of figure~\ref{detunings}, imperfect DIRs at both ends of the array occurs due to the QD detunings. The asymmetric detuning leads to $\mu_{1}^{gg}\neq0$, so that $\ket{\Psi_{f}}_{\rm QDs}\approx\frac{1}{2}(\mu_{1}^{gg}\ket{gg}+\mu_{1}^{gm}\ket{gm}+\mu_{1}^{mg}\ket{mg})$ and there exists a non-zero probability of detection at drain 1 when both the QDs are in $\ket{g}$. Thus, the state $\ket{gg}$ gives a detection, causing a loss of fidelity. Furthermore, in line A we have $\mu_{1}^{gm} = - (\mu_{1}^{mg})^{*}$. This is not a problem if the respective imaginary terms are approximately zero. However, if this is not the case then it leads to dephasing in the final QD state. Here, the phase difference between $\mu_{1}^{gm}$ and $\mu_{1}^{mg}$ affects the relative phase $\phi$ of the generated entangled QD state as $\frac{1}{\sqrt{2}}(\ket{gm}+e^{i\phi}\ket{mg})$. From this point-of-view, $\mu_{1}^{gm} = - \mu_{1}^{mg}$ is highly desirable for achieving a high-fidelity, as pointed out in the limiting scenario in section 3. Thus, increasing $|\Delta\delta/g_{\rm np}|$ causes a loss of fidelity since the respective imaginary terms of $\mu_{1}^{gm}$ and $\mu_{1}^{mg}$ are increased, as can be seen in figure \ref{detunings}~(d). 

In line B of figure~\ref{detunings}, the equal detunings of the QDs enables $\mu_{1}^{gg}=0$ to be obtained via destructive interference with the help of the matching condition, {\it i.e.} $\ket{\Psi_{f}}_{\rm QDs}\approx\frac{1}{2}(\mu_{1}^{gm}\ket{gm}+\mu_{1}^{mg}\ket{mg})$. It also enables $\mu_{1}^{gm} = - \mu_{1}^{mg}$ to be obtained, which provides a high fidelity, in contrast to line A. Here, increasing $|\delta_{0}/g_{\rm np}|$ disturbs the DIRs at both ends of the array, so that the amplitudes related to the occurrence of DIR are no longer much different from the amplitudes that follow the characteristics of the system in the absence of the QDs. Thus, all the other field modes for $\ket{gm}$ are approximately equal to those for $\ket{mg}$ and can be factored out so that $\ket{\Psi_{f}} \rightarrow (\mu_{1}^{gm}\ket{gm}+\mu_{1}^{mg}\ket{mg})\otimes\ket{\rm all~fields}$. In addition, the magnitudes of $\mu_{1}^{gm}$ and $\mu_{1}^{mg}$ decrease with increasing $|\delta_{0}/g_{\rm np}|$ such that $\ket{\Psi_{f}}$ moves closer to the state $\ket{\Psi_{f}}_{\rm QDs}$ obtained in the limiting scenario of weak coherent states. As a result, the fidelities increase with increasing $|\delta_{0}/g_{\rm np}|$ for both $n=2$ and $n=3$, as shown in figure~\ref{detunings}~(e).


\end{document}